\documentclass[journal,twoside,web]{ieeecolor}

\usepackage{generic}
\usepackage{cite}
\usepackage{amsmath,amssymb,amsfonts}
\usepackage{graphicx}
\usepackage{algorithm}
\usepackage{algpseudocode}
\usepackage{textcomp}
\usepackage{multirow} 
\usepackage{multicol} 
\usepackage{threeparttable} 
\usepackage[colorlinks=true, allcolors=blue]{hyperref} 
\usepackage{footnote}
\usepackage{cleveref} 
\usepackage{amssymb}
\usepackage{bbding}
\usepackage{booktabs}
\usepackage{array}

\usepackage[switch]{lineno}

\def\BibTeX{{\rm B\kern-.05em{\sc i\kern-.025em b}\kern-.08em
    T\kern-.1667em\lower.7ex\hbox{E}\kern-.125emX}}
\begin{document}

\title{Pathology-Guided AI System for Accurate Segmentation and Diagnosis of Cervical Spondylosis}
\author{
Qi Zhang\textsuperscript{1}, Xiuyuan Chen\textsuperscript{1}, Ziyi He\textsuperscript{1}, Lianming Wu, Kun Wang\textsuperscript{*}, Jianqi Sun\textsuperscript{*}, and Hongxing Shen\textsuperscript{*}%
\thanks{\textsuperscript{1}Qi Zhang, Xiuyuan Chen, and Ziyi He contributed equally to this work and should be regarded as co-first authors.}%
\thanks{\textsuperscript{*}Corresponding authors: Kun Wang (e-mail: wangkun@renji.com), Jianqi Sun (e-mail: milesun@sjtu.edu.cn), and Hongxing Shen (e-mail: shenhongxing@renji.com).}%
\thanks{This research was supported by the National Natural Science Foundation of China (Grant No.62471297 and No.82272476), 2022 Shanghai Leading Project of the Oriental Talents Program, the National Key Research and Development Program (Grant No.2023YFC2411401), the Fundamental Research Funds for the Central Universities (Grant No.YG2025ZD03), and Clinical Postdoctoral Program of Shanghai Jiao Tong University School of Medicine and Talent Cultivation Project of Renji Hospital (Grant No.RJTJ25-QN-031).}%
\thanks{Qi Zhang and Jianqi Sun are with the School of Biomedical Engineering, Shanghai Jiao Tong University, Shanghai, China.}%
\thanks{Jianqi Sun is also with the National Engineering Research Center of Advanced Magnetic Resonance Technologies for Diagnosis and Therapy (NERC-AMRT), and Med-X Research Institute, Shanghai Jiao Tong University, Shanghai, China.}%
\thanks{Xiuyuan Chen, Kun Wang, and Hongxing Shen are with the Department of Spine Surgery, Renji Hospital, School of Medicine, Shanghai Jiao Tong University, Shanghai, China.}%
\thanks{Ziyi He is with the Department of Radiology, Ruijin Hospital, Shanghai Jiao Tong University School of Medicine, Shanghai, China.}%
\thanks{Lianming Wu is with the Department of Radiology, Renji Hospital, School of Medicine, Shanghai Jiao Tong University, Shanghai, China.}%
}

\maketitle

\begin{abstract}
Cervical spondylosis, a complex and prevalent condition, demands precise and efficient diagnostic techniques for accurate assessment. While MRI offers detailed visualization of cervical spine anatomy, manual interpretation remains labor-intensive and prone to error. To address this, we developed an innovative AI-assisted Expert-based Diagnosis System\footnote{Our code is available at: \url{https://github.com/zhibaishouheilab/CervAI-X}} that automates both segmentation and diagnosis of cervical spondylosis using MRI. Leveraging multi-center datasets of cervical MRI images from patients with cervical spondylosis, our system features a pathology-guided segmentation model capable of accurately segmenting key cervical anatomical structures. The segmentation is followed by an expert-based diagnostic framework that automates the calculation of critical clinical indicators. Our segmentation model achieved an impressive average Dice coefficient exceeding 0.90 across four cervical spinal anatomies and demonstrated enhanced accuracy in herniation areas. Diagnostic evaluation further showcased the system’s precision, with the lowest mean average errors (MAE) for the C2-C7 Cobb angle and the Maximum Spinal Cord Compression (MSCC) coefficient. In addition, our method delivered high accuracy, precision, recall, and F1 scores in herniation localization, K-line status assessment, T2 hyperintensity detection, and Kang grading. Comparative analysis and external validation demonstrate that our system outperforms existing methods, establishing a new benchmark for segmentation and diagnostic tasks for cervical spondylosis.
\end{abstract}


\section{INTRODUCTION}
\label{sec:introduction}
\IEEEPARstart{C}{ervical} spondylosis, a common clinical disease with a prevalence rate rangding from 3.8\% to 17.6\%\cite{nouri2015degenerative}, poses significant diagnostic challenges. Magnetic resonance imaging (MRI) of the cervical spine is radiation-free, multiparametric, and has high soft-tissue resolution\cite{malhotra2017utility}. Due to distinct display of intervertebral discs and spinal cord compression, cervical spine MRI has become the ideal diagnostic evaluation for cervical spondylosis\cite{theodore2020degenerative}. Given the numerous and complex anatomical structures of the cervical spine, the MRI images requires experienced radiologists to achieve the accurate interpretation\cite{malhotra2017utility}. In primary hospitals and emergency settings, non-specialists are more prone to missed and delayed diagnoses. The literature reports that the incidence of delayed diagnosis of cervical spine injuries ranges from 5\% to 20\%\cite{platzer2006delayed}. Therefore, the accurate diagnosis for cervical spondylosis remains a significant clinical challenge.

Recent advances in deep learning have shown promise in improving image analysis and diagnostic precision for cervical spondylosis\cite{park2022assessment}. Deep learning techniques have been applied in areas such as surgical guidance and diagnostic evaluation. Studies have utilized algorithms like U-Net for precise segmentation of cervical vertebrae and intervertebral discs across imaging types, including X-ray, MRI, and CT scans\cite{nozawa2023magnetic,li2021multi,song2023ai,wang2019fully,chen2021study,michopoulou2009atlas}.

However, three main challenges hinder the development of fully automated diagnostic systems for cervical spondylosis. First, existing methods often focus on either segmentation of cervical spine anatomies from MRI or direct diagnosis as detection/classification tasks. Segmentation models may achieve high performance in certain evaluation metrics but often fail to accurately predict pathological areas, leading to poor diagnostic performance. End-to-end detection and classification tasks also lack the ability to quantify disease severity, reducing their clinical interpretability\cite{teng2022survey,vellido2020importance}. Some studies have explored AI-assisted diagnostic systems that follow a segmentation-to-diagnosis workflow, but significant gaps exist between these two steps, limiting the clinical applicability\cite{al2024advanced}. 
Additionally, the lack of large, well-annotated datasets further hinders system reliability and effectiveness.

Second, cervical spondylosis is a complex disease encompassing multiple pathological conditions, such as disc herniation, spinal cord compression, spinal stenosis, and T2 hyperintensity spinal cord lesions. A comprehensive diagnostic framework incorporating various clinical indicators and parameters is essential for thorough disease evaluation. However, most existing diagnostic methods focus on limited indicators, reducing their ability to provide a comprehensive assessment of disease severity for clinicians and patients\cite{merali2021deep,ma2020faster,zhong2024sagittal}. The absence of a comprehensive set of diagnostic indicators prevents these systems from generating holistic reports similar to those of expert clinicians.

Third, expert involvement is crucial in the development of AI-assisted diagnostic systems. Experts contribute in three key areas: 1) guiding the selection of clinically significant indicators and optimal parameters, 2) proposing novel methods and indicators based on data-driven insights and clinical observations, and 3) validating the system’s diagnostic performance through comprehensive evaluations. Our ultimate goal is to achieve expert-level diagnostic precision, enabling the AI system to emulate expert decision-making. For instance, incorporating clinical knowledge and expert observations as prior knowledge can direct the model’s attention to key regions and improve diagnostic accuracy. Nevertheless, few studies fully recognize the importance of expert knowledge, resulting in its limited integration throughout the system development process.

\begin{figure}[!ht]
\centering
\includegraphics[width=\columnwidth]{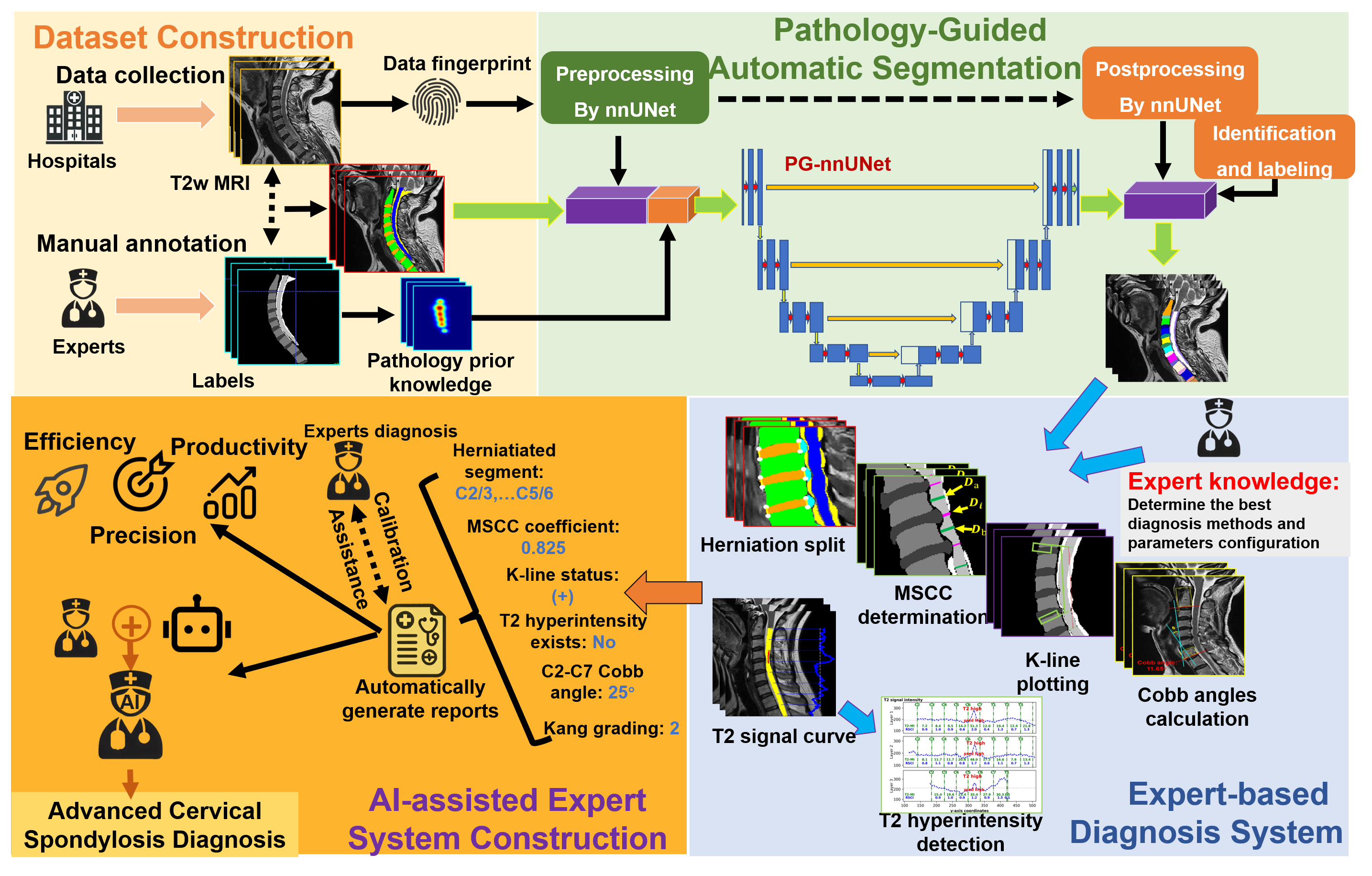}
\caption{\label{fig1} Overall workflow of our AI-assisted Expert-based Diagnosis System for cervical spondylosis.}
\end{figure}

To address these challenges, this study proposes an AI-assisted Expert-based Diagnosis System for cervical spondylosis. The system integrates pathology-guided automatic segmentation and an expert-based diagnostic framework. We developed a segmentation framework based on nnUNet, referred to as Pathology-Guided nnUNet (PG-nnUNet), and incorporated an auxiliary task to detect pathological areas. Following segmentation, an expert-based diagnosis system was constructed using the segmentation results, incorporating a comprehensive framework based on six essential clinical indicators derived from clinical parameters and expert knowledge. Additionally, we introduce a novel parameter to improve T2 hyperintensity detection. Finally, we evaluated the system’s diagnostic results through expert measurement comparison, demonstrating the effectiveness of our automated approach. The overall workflow is illustrated in \Cref{fig1}.

The main contributions of our study are:
\begin{itemize}
\item Proposal of an AI-assisted Expert-based Diagnosis System for cervical spondylosis, achieving automation and precision in both segmentation and diagnosis using T2-weighted cervical MRI.
\item Introduction of a PG-nnUNet segmentation network based on nnUNet, incorporating pathology prior knowledge and multi-task learning to enhance pathological area segmentation accuracy.
\item Development of a multi-parameter diagnostic framework based on segmentation results and expert knowledge, offering a more comprehensive and reliable diagnosis than existing methods.
\item PG-nnUNet achieved state-of-the-art segmentation performance across multiple cervical spinal anatomies, validated by expert measurements to meet clinical requirements and outperform other approaches.
\end{itemize}

\section{Related Works}
\subsection{Segmentation Methods for Cervical Spine}
Deep learning has driven the development of advanced segmentation models for the cervical spine. Zhang et al.\cite{zhang2023fully} enhanced $U^{2}-Net$ for automatic cervical vertebrae segmentation in CT images, while Rehman et al.\cite{rehman2019novel} introduced a U-Net-based method for X-ray images. Although CT and X-ray are widely used, MRI is superior in soft-tissue resolution, making it critical for clinical decision-making. Several studies have explored spinal MRI segmentation. Xhelili et al.\cite{xhelili2023automatic} compared three deep learning methods for vertebrae segmentation in MR volumes, suggesting the potential of adapting CT-based models to MRI tasks. Huang et al.\cite{huang2020spine} presented Spine Explorer, which uses U-Net to segment lumbar MRIs and quantify spinal structures. However, most studies focus on lumbar or thoracic regions rather than cervical MRI, especially pathological cases. Furthermore, they often overlook the impact of segmentation results on disease diagnosis, limiting clinical utility.

Cervical spine segmentation faces key challenges: (1) the complex anatomical structure, particularly the variable shapes of the spinal cord and cerebrospinal fluid\cite{andrew2020spine}; (2) limited availability of cervical MRI data compared to X-ray or CT\cite{zhang2022seuneter}; and (3) difficulty in accurately segmenting pathological areas crucial for diagnosis\cite{bueno2022automated,graf2023denoising,bogduk2016functional}. To address these, we enhanced nnUNet, known for its self-configuration capabilities in medical image segmentation\cite{isensee2021nnu}. By incorporating prior knowledge into nnUNet and using a T2-weighted cervical MRI dataset from patients with disc herniation, we improved pathological area segmentation. This study not only boosts segmentation accuracy but also bridges the gap between segmentation quality and automated disease diagnosis, underscoring segmentation's essential role in clinical workflows.

\subsection{Diagnosis Systems for Cervical Spondylosis}
Current diagnosis methods for cervical spondylosis can be categorized into two main types. The first type directly predicts a single or a few clinical indicators, focusing on specific pathologies. For instance, Merali et al.\cite{merali2021deep} developed a ResNet-50-based classification model to categorize spinal cord compression using axial T2-weighted MRI images. Ma et al.\cite{ma2020faster} utilized Faster-RCNN to detect cervical spinal cord injury and disc degeneration in MRI scans. Zhong et al.\cite{zhong2024sagittal} measured sagittal balance parameters on cervical spine T2 MR images via superpixel segmentation. While these end-to-end models offer task-specific accuracy, they often fail to provide a comprehensive evaluation of cervical spondylosis. Moreover, the ``black-box'' nature of deep learning models hinders interpretability, reducing clinical trust and applicability.

The second type integrates segmentation with subsequent clinical indicator calculations. Hohenhaus et al.\cite{hohenhaus2024quantification} segmented spinal cord and cerebrospinal fluid in 3D MRI to quantify cervical spinal stenosis. Shawwa et al.\cite{al2024advanced} employed a specialized toolbox to segment high-resolution MRI scans and measure metrics like cord diameter and length, predicting degenerative cervical myelopathy severity. However, the effectiveness of these systems is often constrained by insufficient segmentation precision and robustness. Inaccurate segmentation degrades clinical indicator accuracy, compromising diagnostic reliability and hindering clinical adoption.

We propose a precise, automated diagnostic framework for cervical spondylosis. Our pathology-guided segmentation method outperforms existing models, particularly in pathological regions. Leveraging these segmentations, we develop an expert knowledge-based diagnostic system integrating multiple quantitative clinical indicators. This approach overcomes limitations of prior methods, delivering a more comprehensive and reliable solution.

\section{Material and methods}

\subsection{Data collection and annotation}

\begin{table}[ht]
\centering
\caption{Characteristics of the Renji and Ruijin Dataset}
\label{tab:datasets}
\begin{tabular}{@{}lcc@{}}
\toprule
\textbf{Category} & \textbf{Renji dataset} & \textbf{Ruijin dataset} \\ \midrule
\multicolumn{3}{l}{\textbf{Patients}} \\ \midrule
No. of patients & 952 & 95 \\
No. of women & 434 (45.6\%) & 45 (47.3\%) \\
Age (yr*) & 60.8±10.8 & 63.0±10.0 \\ \midrule
\multicolumn{3}{l}{\textbf{Imaging}} \\ \midrule
No. of scans & 960 & 95 \\
No. of V (C2-C7) & 5760 & 570 \\
No. of IVD (C2/C3-C6/C7) & 4800 & 475 \\ \midrule
\multicolumn{3}{l}{\textbf{Kang grading}} \\ \midrule
Grade 1 (mild, \% of total) & 220 (22.9\%) & 50 (52.6\%) \\
Grade 2 (moderate, \% of total) & 347 (36.1\%) & 37 (38.9\%) \\
Grade 3 (severe, \% of total) & 393 (40.9\%) & 8 (8.4\%) \\ \bottomrule
\multicolumn{3}{l}{* Data are means ± standard deviations} \\
\multicolumn{3}{l}{V: vertebral bodies; IVD: intervertebral discs} \\
\multicolumn{3}{l}{\textbf{Ruijin dataset} is used as an external validation set} \\
\end{tabular}
\end{table}
This study was granted exemption by the ethics committee of Renji Hospital and Ruijin Hospital (Approval No: RA-2025-215). We certify that the study was performed in accordance with the 1964 declaration of HELSINKI and later amendments. Patients' cervical MR T2-weighted images were collected from the Picture Archiving and Communication System (PACS) to construct the datasets. {RENJI dataset was acquired using United Imaging 3.0T and GE 3.0T scanners, while the RUIJIN dataset was collected with United Imaging 3.0T scanners. To ensure data consistency across different devices, we standardized the MRI acquisition protocols for all scanners. Specifically, all scans followed identical T2-weighted protocols: TR/TE = 1800/112.8 ms, slice thickness = 3 mm, matrix = 384×384, FOV = 220×220 mm. \Cref{tab:datasets} summarizes the key characteristics of the two datasets. The degree of cervical spine degeneration was assessed using the Kang grading\cite{lee2020inter} based on midsagittal MRI. Grade 0: no stenosis; Grade 1: mild stenosis, subarachnoid occlusion \textless 50\%; Grade 2: moderate stenosis, central spinal canal stenosis with compression and deformation of the spinal cord, but no signal change in the spinal cord; Grade 3: severe stenosis, stenosis with spinal cord T2 hyperintensity. Each image was a sagittal single-slice sequence of the cervical spine with a resolution of 512 × 512. Four regions—vertebral bodies (V), intervertebral discs (IVD), spinal cord (SC), and cerebrospinal fluid (CSF)—were manually annotated by a senior spine surgeon with over 10 years of experience. The Renji dataset was split into training (672 cases) and internal test sets (288 cases) at a 7:3 ratio, while the Ruijin dataset (95 cases) served as the external validation set.

\subsection{Pathology-Guided Cervical Spine Segmentation with PG-nnUNet}
Our cervical spine segmentation approach follows a three-step process: data preparation, network training, and inference, as illustrated in \Cref{fig2}.

\begin{figure}[!ht]
\centering
\includegraphics[width=\columnwidth]{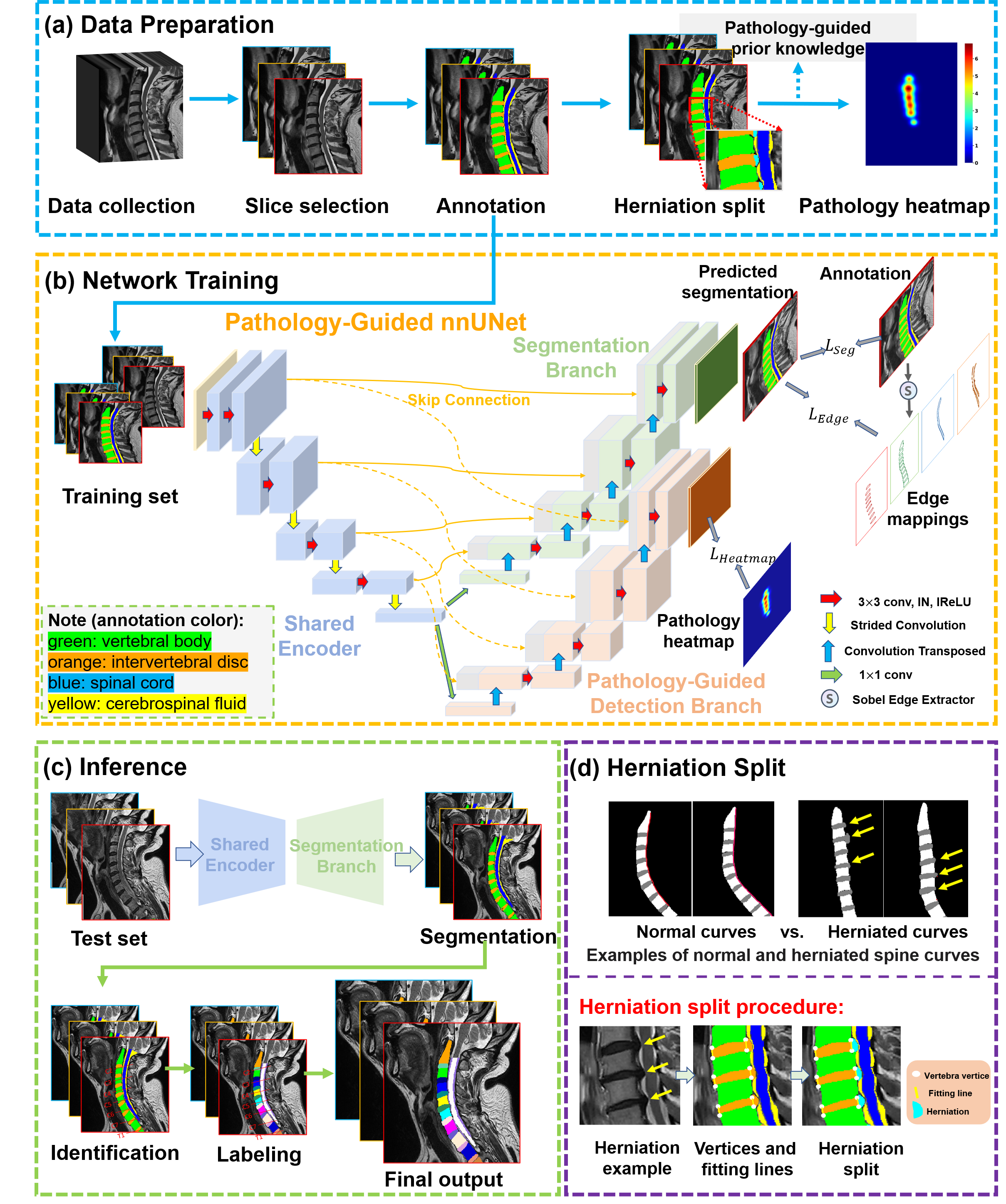}
\caption{\label{fig2} Comprehensive workflow for cervical spine MRI segmentation. (a) Data preparation, including annotation and pathology heatmap generation. (b) PG-nnUNet architecture designed for accurate segmentation of multiple anatomical structures and pathological areas. (c) Inference and post-processing for subsequent analysis. (d) Procedure for disc herniation segmentation.}
\end{figure}

\subsubsection{Data preparation}\label{section:B}
In clinical diagnosis, experts focus on the central three sagittal slices of cervical MRI sequences for critical anatomical details. These slices were selected from each MRI sequence in the training and test sets. The four target regions (V, IVD, CSF, SC) were manually annotated using ITK-SNAP\cite{py06nimg}. To capture additional anatomical content and prevent under-segmentation, regions below the C6/7 intervertebral disc were also included in the segmentation task.

After annotating the primary anatomical structures, we segment disc herniations as illustrated in \Cref{fig2}(d). Normal spinal anatomy features near-linear, smooth edge contours, while herniations appear as protrusions deviating from this pattern (denoted by yellow arrows). Contour points of vertebral bodies were extracted, and connecting lines between adjacent vertebrae defined the normal intervertebral disc area. Disc protrusions beyond these lines were classified as herniations and highlighted in sky blue. These herniation segmentations, though not directly used in network training, served as prior knowledge to guide the model toward pathological areas. Pathology heatmaps were then generated to represent these herniations for further diagnostic processing.

\subsubsection{Network training}
To achieve accurate segmentation of multiple cervical spine regions, particularly pathological areas, we propose the Pathology-Guided nnUNet (PG-nnUNet). Built upon the 2D nnUNet baseline, this framework enhances diagnostic accuracy by incorporating a pathology-guided detection branch. This branch prioritizes critical pathological regions, ensuring the model focuses on areas vital for diagnosis.

\subsubsection{Pathology heatmap generation}\label{section:A}
We convert herniation areas into a pathology heatmap used as supervision or prior knowledge in the PG-nnUNet. \Cref{fig3} and \Cref{alg1} detail this process. First, connected components of herniation segmentations are extracted, discarding those smaller than $Size_{min}$. For each herniation component $R_i$, we compute its centroid $(x_{\mu}^i, y_{\mu}^i)$ and size $|R_i|$. Using a Gaussian distribution, we generate a heatmap $H_i(x, y)$ for each component:
\begin{equation}\label{eq1}
\begin{aligned}
    H_i(x, y) = \exp\left(-\frac{(x - x_{\mu}^i)^2 + (y - y_{\mu}^i)^2}{2\sigma_i^2}\right) \times \sqrt{|R_i|}
\end{aligned}
\end{equation}
The standard deviation $\sigma_i$ controls the spread of the herniation region and is related to the herniation size: $\sigma_i = \sigma_{\text{scale}} \times \sqrt{|R_i|}$. The larger the herniation, the greater the severity of the pathology, $\sqrt{|R_i|}$ is used as a scaling factor to adjust the intensity at the herniation center. Adding all individual heatmaps pixel-wise creates the overall pathology heatmap $H(x, y)$, where pixel intensity reflects pathology likelihood and severity, with higher intensities indicating regions of greater pathological influence and lower intensities indicating less.

\begin{figure}[!ht]
\centering
\includegraphics[width=\columnwidth]{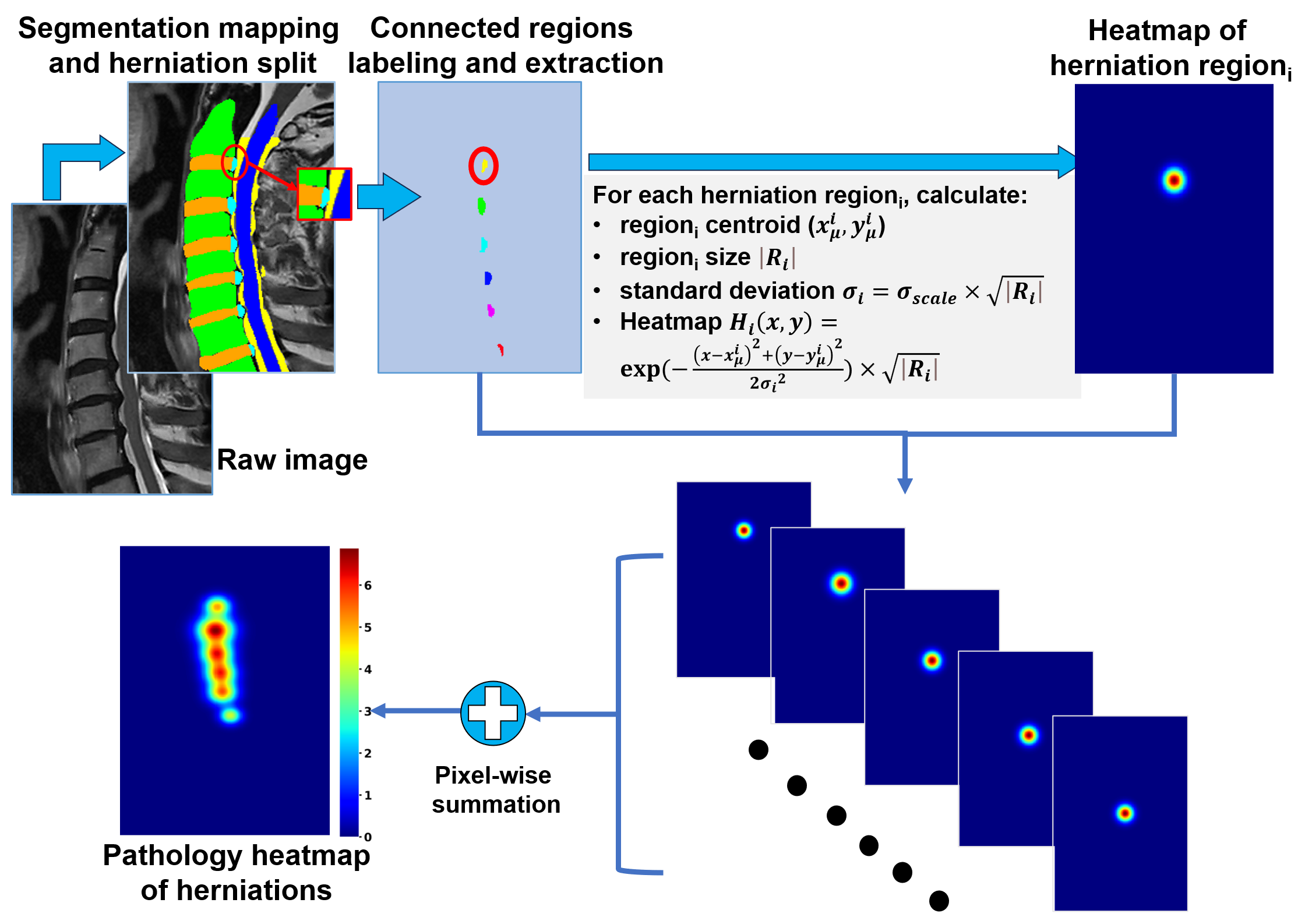}
\caption{\label{fig3} The generation of each herniation heatmap and the overall pathology heatmap. A higher intensity in the heatmap indicates a stronger pathological influence.}
\end{figure}

\begin{algorithm}[!ht]
\caption{Pathology Heatmap Generation Algorithm}\label{alg1}
\begin{algorithmic}
\Require Segmentation map of herniation $S$, hyperparameters $\sigma_{\text{scale}} = 1.5$, $\text{min\_region\_size} = 10$
\Ensure Generated pathology heatmap $H(x, y)$

\State Apply connected component labeling on $S$ to identify all connected regions $R = \{R_1, R_2, \dots, R_n\}$, where each $R_i$ is a binary mask representing a connected region.
\For{each region $R_i$ in $R$}
    \If{region size $|R_i| < \text{min\_region\_size}$}
        \State Continue to the next region
    \EndIf
    \State Compute the centroid $(x_{\mu}^i, y_{\mu}^i)$ of region $R_i$
    \State Calculate the region size $|R_i|$
    \State Compute the standard deviation $\sigma_i$:
    \[
    \sigma_i = \sigma_{\text{scale}} \times \sqrt{|R_i|}
    \]
    \State Generate a Gaussian heatmap $H_i(x, y)$ centered at $(x_{\mu}^i, y_{\mu}^i)$ with standard deviation $\sigma_i$:
    \[
    H_i(x, y) = \exp\left(-\frac{(x - x_{\mu}^i)^2 + (y - y_{\mu}^i)^2}{2\sigma_i^2}\right) \times \sqrt{|R_i|}
    \]
\EndFor
\State Combine all generated heatmaps $H_i(x, y)$ by summing them pixel-wise to form the overall pathology heatmap $H(x, y) = \sum_i H_i(x, y)$

\Return $H(x, y)$
\end{algorithmic}
\end{algorithm}

\subsubsection{Network architecture}
The PG-nnUNet employs a multi-task learning framework built upon the original nnUNet's encoder-decoder architecture. This framework trains the model to perform two related tasks simultaneously: 1) generating segmentation probability maps for the key anatomical structures (V, IVD, SC, CSF), and 2) predicting pathology heatmaps highlighting regions likely to contain disc herniations. PG-nnUNet utilizes a shared encoder and a dual-branch decoder. Images are input into the encoder, where features are extracted through convolution and max-pooling layers. These features, containing both high- and low-level information, are passed to the decoder and through a softmax layer to produce the segmentation map. Each encoder layer is linked to a corresponding decoder layer via skip-connections, enabling the integration of local details and global context. Beyond the original segmentation decoder, we introduce a Pathology-Guided Detection Branch post-encoder, mirroring the Segmentation Branch's architecture.


Unlike object detection heatmaps, which focus on fixed objects and peak intensity points, our pathology heatmap strategy considers the size, location, and quantity of pathological herniations. This approach not only aids detection but also enhances segmentation by emphasizing pathological regions challenging to segment due to their small size and similarity to adjacent normal tissues. To optimize the network, we apply an $\mathcal{L}_1$ loss between the predicted and true heatmaps:
\begin{equation}\label{eq2}
\begin{aligned}
    \mathcal{L}_{Heatmap}=||H_{pred}-H_{true}||_1
\end{aligned}
\end{equation}
Where $H_{pred}$ and $H_{true}$ are the predicted and true heatmaps, respectively, and pixel intensity reflects pathological influence. For segmentation, we use a combined loss function of cross-entropy and Dice loss:
\begin{equation}\label{eq3}
\begin{aligned}
    \mathcal{L}_{Seg}={L}_{CE}+{L}_{Dice}
\end{aligned}
\end{equation}
Given the importance of edge details, particularly in ambiguous or pathological regions, we generate distinct edge maps for the four target anatomies using the Sobel function. The predicted edge maps are obtained from the predicted segmentation maps, and $\mathcal{L}_1$ loss between the predicted and ground truth edge maps is calculated:
\begin{equation}\label{eq4}
\begin{aligned}
    \mathcal{L}_{Edge}&=0.25*(||Sobel(S_{pred}^{IVD})-Sobel(S_{GT}^{IVD})||_1\\
    &+||Sobel(S_{pred}^{V})-Sobel(S_{GT}^{V})||_1\\
    &+||Sobel(S_{pred}^{SC})-Sobel(S_{GT}^{SC})||_1\\
    &+||Sobel(S_{pred}^{CSF})-Sobel(S_{GT}^{CSF})||_1)
\end{aligned}
\end{equation}
Where $S^{IVD}, S^{V}, S^{SC}, S^{CSF}$ denote the binary segmentation maps for the respective regions, and $Sobel(\cdot)$ applies the Sobel function to extract the edge maps. The total loss is:
\begin{equation}\label{eq5}
\begin{aligned}
    \mathcal{L}_{total}=\mathcal{L}_{Seg}+\lambda_{h}\cdot\mathcal{L}_{Heatmap}+\lambda_{e}\cdot\mathcal{L}_{Edge}
\end{aligned}
\end{equation}

Here, $\lambda_{h}$ and $\lambda_{e}$ are the hyper-parameters that control the emphasis on pathology heatmap and edge losses.

\subsubsection{Inference}
The Pathology-Guided Detection Branch is removed during inference, and the network outputs predicted segmentation maps for test images. Beyond standard nnUNet post-processing, we implement a procedure to identify and label each cervical spine segment, critical for subsequent diagnosis. Structures are sorted along the Y-axis, with the model selecting six contiguous connected regions for vertebrae (C2 to C7) and five for intervertebral discs (C2/3 to C6/7). Each structure is labeled numerically: 1 to 6 for vertebrae and 7 to 11 for intervertebral discs, facilitating further analysis and diagnosis.

\subsection{Expert-Knowledge based Diagnosis System Development}\label{section:D}
Using the cervical spine segmentation from our study, we developed an Expert-Knowledge based Diagnosis System to address spinal pathologies like disc herniation and spinal stenosis. This system automates the computation of key clinical indicators, including the MSCC coefficient, T2 hyperintensity detection within the spinal cord, and Cobb angle measurement. Based on these calculated metrics, Kang grading is achieved. The system's framework is shown in \Cref{fig4}.

\subsection{MSCC Processing and Calculation}
The Maximum Spinal Cord Compression (MSCC) coefficient\cite{fehlings1999optimal} is a crucial indicator in assessing the severity of disc herniation and spinal cord compression. It compares the spinal canal width in affected and unaffected areas, with higher values indicating more severe compression.

After segmenting and localizing disc herniations, the system calculates the MSCC coefficient using:
\begin{equation}\label{eq6}
\begin{aligned}
    \text{MSCC (\%)} = \left(1 - \frac{D_i}{\left(\frac{D_a + D_b}{2}\right)}\right) \times 100\%
\end{aligned}
\end{equation}
Here, $D_i$ is the narrowest diameter of the $i_{th}$ spinal canal at the herniation level. $D_a$ and $D_b$ are the diameters of the normal spinal canal segments located above and below the herniation, respectively, as illustrated in \Cref{fig4}(a). The MSCC coefficient quantifies spinal canal compression, with values near 100\% indicating significant compression and values near 0\% suggesting a less severe condition. \Cref{fig4}(b) shows an example of MSCC calculation results.

\begin{figure}[!ht]
\centering
\includegraphics[width=\columnwidth]{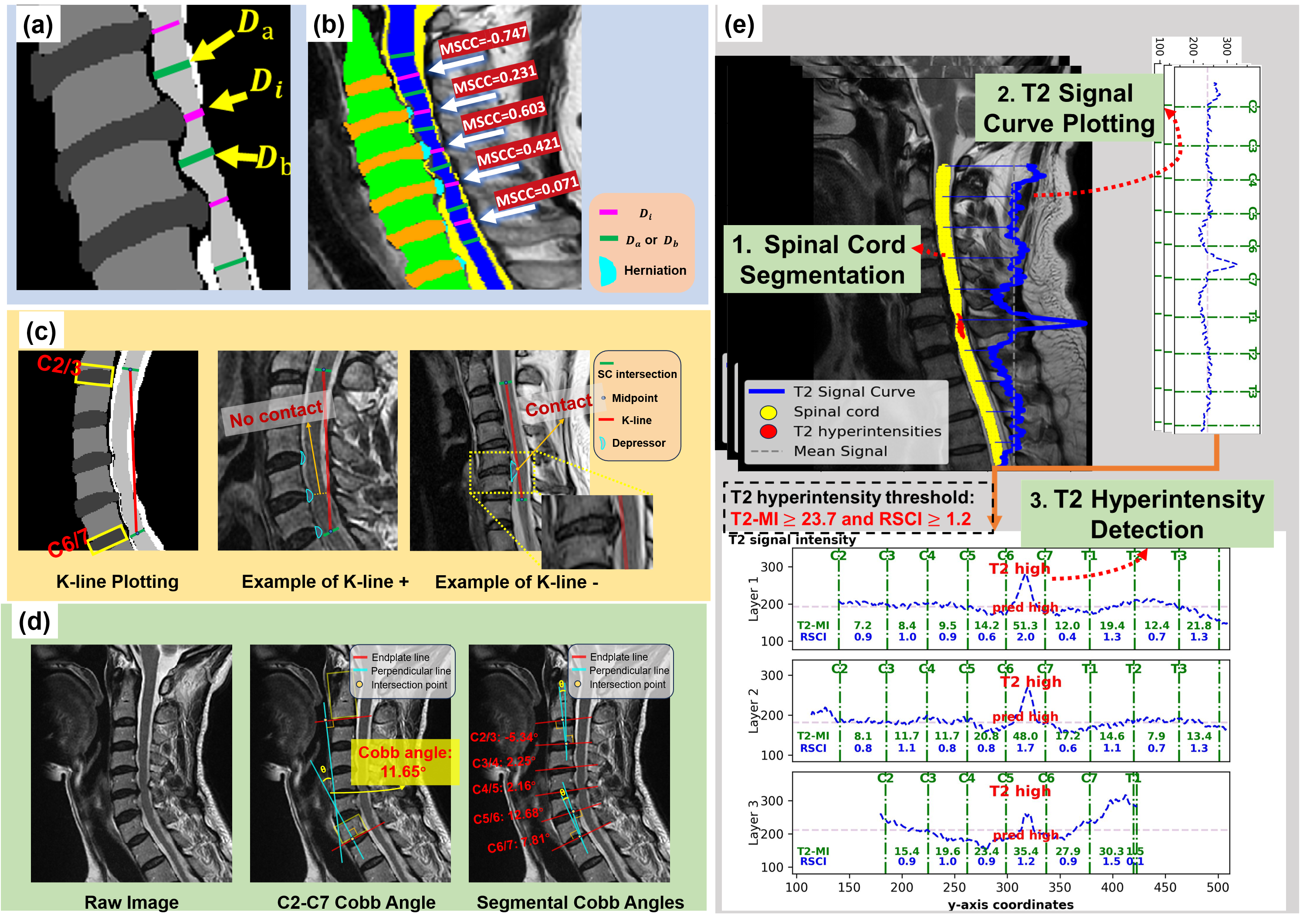}
\caption{\label{fig4} Framework of the Expert-Knowledge Based Diagnosis System and calculation methods. (a) Measuring spinal cord widths; (b) MSCC calculation example; (c) Modified K-Line drawing; (d) Cobb angle calculation; (e) T2 hyperintensity detection. 'T2 high' and 'pred high' indicate actual and predicted T2 hyperintensities.}
\end{figure}

\subsection{Modified K-Line Drawing Process}\label{section:C}
The modified K-line\cite{taniyama2014modified} connects the spinal cord midpoints at C2 and C7, offering insights into spinal deformities and aiding surgical planning by illustrating the relationship between cervical spine curvature, compressive lesions, and the spinal cord. This process is automated post-segmentation and labeling of intervertebral discs and vertebral bodies, as shown in \Cref{fig4}(c).

\subsubsection{Identification of relevant vertebral bodies}
In the Postprocessing module, the system identifies the second cervical (C2) and seventh cervical (C7) vertebral bodies within the segmentation. For each identified vertebra, a minimal rectangular bounding box (yellow) is extracted.

\subsubsection{Determining spinal cord midpoints}
At the C2 level, the upper edge of the minimal rectangular bounding box is identified and extended to intersect with the spinal cord segmentation. The midpoint of this intersection line is designated as the C2 spinal cord midpoint. A similar procedure is conducted at the C7 vertebra using the lower edge of its bounding box. The Modified K-Line is drawn by connecting the calculated midpoints at C2 and C7.

\subsubsection{K-line assessment(±)}
A K-line positive (K-line +) indicates no contact between the anterior depressor (mainly intervertebral herniation) and the K-Line, while a K-line negative (K-line -) indicates contact. Examples are shown in \Cref{fig4}. These automated assessments are validated by clinical evaluations, confirming the algorithms' accuracy in identifying these critical conditions.

\subsection{Automated Calculation of the Cobb Angle}
The Cobb angle, a key indicator for evaluating spinal deformities such as scoliosis, is automatically determined from the segmentation results.

Minimal bounding boxes are extracted from each vertebral body segment. For C2-C7 Cobb angle calculation, lines are extended along the lower endplates of C2 and C7 bounding boxes. Perpendicular lines are drawn from these extensions, and the intersection angle represents the Cobb angle, visualized in \Cref{fig4}(d). This method also allows for segmental Cobb angle calculation, providing localized spinal curvature assessment. For instance, a negative Cobb angle at C2/3 indicates kyphosis.

\subsection{Detection of T2 hyperintensity}
\label{T2_hyperintensity}
The method consists of three main steps: spinal cord segmentation, T2-SI (T2 signal intensity) curve plotting, and indicators computation.

\subsubsection{Spinal Cord Segmentation}
The spinal cord is segmented based on cervical vertebrae anatomical landmarks, dividing it into specific levels like C2/3 or C5/6.

\subsubsection{T2-SI Curve Plotting}
A  T2-SI curve is generated by calculating the average signal intensity across the spinal cord at each level of y-axis. This curve provides a graphical representation of the distribution of T2 signal values along the length of the spinal cord.

\subsubsection{T2-MI (T2 Myelopathy Index) Calculation}
The T2-MI \cite{hohenhaus2023automated} for each spinal cord segment is calculated using:

\begin{equation}\label{eq7}
\begin{aligned}
    T2-MI = \frac{T2-SI_{\text{range}}}{T2-SI_{\text{mean}}}
\end{aligned}
\end{equation}

Here, T2-SI$_{\text{range}}$ is the signal intensity variation range within the segment, and T2-SI$_{\text{mean}}$ is the average signal intensity. 

\subsubsection{Introduction to Relative Signal Change Index (RSCI)}
To better compare signal intensity changes across spinal segments, we introduce the Relative Signal Change Index (RSCI), defined as:
\begin{equation}\label{eq8}
\begin{aligned}
    RSCI_{i} = \frac{3 \times T2-MI_{i}}{\sum_{k=i-1}^{i+1}(T2-MI_{k})}, \quad 2 \leq i \leq n-1
\end{aligned}
\end{equation}

Unlike T2-MI, which measures absolute signal intensity change within a single segment, RSCI provides a relative measure by comparing a segment's T2-MI to those of adjacent segments.  The T2-MI measures the absolute signal intensity change only within a single spinal segment. In cases where a patient’s MRI shows inherently high variance in signal intensity across the spinal cord, T2-MI values might be uniformly high across all segments. This uniformity can mask the presence of pathologically high signals, as T2-MI does not account for relative changes between adjacent segments.  The pathological segment with high signal intensity will have a significantly higher T2-MI compared to its neighboring segments. The RSCI amplifies this difference, making it easier to identify segments with pathologically high signals.

\subsection{Automated Kang Grading System}
\label{sec:kang_grading}

The Kang grading system was implemented as a rule-based module using our framework's segmentation outputs and quantitative indicators. Grading was performed per intervertebral segment (C2/3 to C6/7) on the mid-sagittal slice, classifying each segment as Kang 0 (normal), 1 (mild), 2 (moderate), or 3 (severe stenosis). This automated pipeline replicates clinical criteria \cite{kang2011new} while ensuring consistency through quantitative geometric and signal-based measurements.
\subsubsection{Herniation Detection}

For each intervertebral segment $i$, herniation presence is determined by: $\mathcal{H}_i=\text{IVD}_i \cap \mathcal{H}$, where $\text{IVD}_i$ is the segmented intervertebral disc, $\mathcal{H}$ is the herniated area segmented in \Cref{section:B}, $\cap $ indicates the intersection area. If $\mathcal{H}_i$ doesn't exist, the segment is classified as Grade 0 (no stenosis).

\subsubsection{Stenosis Severity Assessment}

For segments with $\mathcal{H}_i$, stenosis severity is quantified through the defined stenosis ratio: $\text{Ratio}_i^{stenosis} = (1-\frac{D_i^{\text{hern}}}{\overline{D}_i^{\text{ref}}})\times 100\%$. Where $D_i^{\text{hern}}$ measures the nearest distance between $\mathcal{H}_i$ and spinal cord, and $\overline{D}_i^{\text{ref}}$ is the average of two reference distances: the distance from the midpoint of the superior vertebral body's posterior wall to the spinal cord boundary and the distance from the midpoint of the inferior vertebral body's posterior wall to the spinal cord. Higher $\text{Ratio}_i^{\text{stenosis}}$ indicates more severe stenosis.

\subsubsection{Grade Assignment}
The segmental grade is determined based on geometric relationships and T2 signal characteristics:
\[
\text{Grade}_i = 
\begin{cases} 
3 & \text{if } T2-hyper_i=1 \\
2 & \text{if } D_i^{\text{hern}} = 0 \\
1 & \text{if } \text{Ratio}_i^{stenosis} \geq 50\% \\
0 & \text{if } \text{Ratio}_i^{stenosis} < 50\% \\
\end{cases}
\]

$\text{T2-hyper}_i=1$ indicates the existence of T2 hyperintensity, which is detected using the T2-MI and RSCI framework in \Cref{T2_hyperintensity}. $D_i^{\text{hern}} = 0$ indicates that the contact between $\mathcal{H}_i$ and spinal cord. The final Kang grade is a patient-level grade that reflects the most severe segmental grade.

\section{Experiments}
\subsection{Comparison with Other Segmentation Methods}
To evaluate our model's segmentation performance across multiple cervical spinal anatomies, we compared it with several state-of-the-art models, divided into CNN-based and Transformer-based categories. For CNN-based models, we selected Standard UNet\cite{ronneberger2015u}, Attention UNet (AttUnet)\cite{oktay2018attention}, Nested UNet (UNet++)\cite{zhou2018unet++}, CENet\cite{gu2019net}, SegNet\cite{badrinarayanan2017segnet}, and Probabilistic UNet (ProUNet)\cite{kohl2018probabilistic}. For Transformer-based models, we included Vision Transformer (ViT)\cite{dosovitskiy2020image}, SwinUNet\cite{cao2022swin}, and TransUNet\cite{chen2021transunet}. All models are 2D architectures. We also incorporated Segment Anything Model 2 (SAM 2)\cite{kirillov2023segany}, a popular foundation model for visual segmentation tasks. Additionally, we compared our method with the original 2D nnUNet\cite{isensee2021nnu} (nnUNet baseline) and a variant with edge loss (proposed in \Cref{eq4}) to assess our approach's effectiveness.

\subsection{Implementation Details}
All experiments were implemented in PyTorch on a workstation with an NVIDIA 4090 GPU. To ensure robustness and reduce outlier impact, we used five-fold cross-validation. The training set was divided into five subsets, with each subset validated in turn while the remaining four were used for training. The best model from each run was selected based on validation performance, and the test set was evaluated using these models. Final performance was reported as the average metric values across all runs.

For all segmentation models, we used the Adam optimizer with an initial learning rate of $1e^{-4}$ and a batch size of 12. CNN-based models were trained for 400 epochs, while Transformer-based models were trained for 1000 epochs. Image intensities were normalized to [0,1] using Z-score normalization, and voxel sizes were resampled to [1 mm, 1 mm]. Images were center-cropped to [256, 256] to focus on the region of interest (ROI). Data augmentation, including 2D rotations, scaling, translation, and flipping, was applied to enhance training data diversity. The segmentation loss was a combination of cross-entropy loss and Dice loss, as described in \Cref{eq3}.

For SAM 2, we used the ViT-H SAM version with pre-trained checkpoints. SAM 2 requires prompts like points or boxes. In our experiments, a senior spine surgeon manually drew rough boxes around objects of interest, which were used as prompts for SAM 2 to generate segmentation masks for comparison.

For ViT-based segmentation, we modified the original Masked Autoencoder (MAE)\cite{he2022masked} by adding a lightweight decoder for pixel-level segmentation. The ViT encoder's output patch embeddings were passed through a decoder with 8 transformer blocks and a transpose convolution layer for upsampling. The encoder was pre-trained using a self-supervised MAE framework on the ImageNet-21K dataset.

For nnUNet-based models, we adopted the default preprocessing and post-processing modules provided by nnUNet. Pathology heatmaps were preprocessed to match the image size and data augmentation settings. The training epoch was set to 300 to ensure convergence. To systematically evaluate the contributions of different components, we implemented three nnUNet variants. (1) nnUNet baseline: the original 2D nnUNet trained with the combined cross-entropy and Dice loss ($\mathcal{L}_{Seg}$, \Cref{eq3}). (2) nnUNet+Edge loss: the baseline nnUNet augmented with the edge loss term ($\mathcal{L}_{Edge}$, \Cref{eq4}). Its total loss is $\mathcal{L}_{total}=\mathcal{L}_{Seg}+\lambda_{e}\cdot\mathcal{L}_{Edge}$, with $\lambda_{e}=1$. (3) PG-nnUNet: the proposed model incorporating both the Pathology-Guided Detection Branch (supervised by $\mathcal{L}_{Heatmap}$, \Cref{eq2}) and the edge loss. Its total loss is $\mathcal{L}_{total}=\mathcal{L}_{Seg}+\lambda_{h}\cdot\mathcal{L}_{Heatmap}+\lambda_{e}\cdot\mathcal{L}_{Edge}$. Comprehensive sensitivity analysis (\Cref{tab:hyperparam}) showed $\lambda_h=1$ and $\lambda_e=1$ achieved optimal balance, yielding the highest herniation Dice score (0.6599), which is clinically critical for pathological segmentation. While $\lambda_e=1.5$ gave marginally better macro-average Dice (0.8711 vs 0.8707), it reduced herniation performance (0.6572). We therefore selected $\lambda_h=\lambda_e=1$ to prioritize herniation detection while maintaining excellent overall performance.

\begin{table}[ht]
\centering
\caption{Hyperparameter sensitivity analysis for $\lambda_h$ and $\lambda_e$}
\label{tab:hyperparam}
\resizebox{\columnwidth}{!}{
\begin{tabular}{cccccc}
\toprule
\multicolumn{3}{c}{$\lambda_e=1$} & \multicolumn{3}{c}{$\lambda_h=1$} \\
\cmidrule(r){1-3} \cmidrule(l){4-6}
$\lambda_h$ & \begin{tabular}{@{}c@{}}Average Dice\\ (Macro)\end{tabular} & \begin{tabular}{@{}c@{}}Herniation\\ Dice\end{tabular} &
$\lambda_e$ & \begin{tabular}{@{}c@{}}Average Dice\\ (Macro)\end{tabular} & \begin{tabular}{@{}c@{}}Herniation\\ Dice\end{tabular} \\
\midrule
0 & 0.8669 & 0.6469 & 0 & 0.8624 & 0.6554 \\
0.5 & 0.8701 & 0.6588 & 0.5 & 0.8677 & 0.6581 \\
\textbf{1} & \textbf{0.8707} & \textbf{0.6599} & \textbf{1} & 0.8707 & \textbf{0.6599} \\
1.5 & 0.8698 & 0.6592 & 1.5 & \textbf{0.8711} & 0.6572 \\
2 & 0.8676 & 0.6576 & 2 & 0.8673 & 0.6548 \\
\bottomrule
\multicolumn{6}{p{\linewidth}}{Average Dice: macro average Dice score of five regions, including intervertebral discs, vertebral bodies, cerebrospinal fluid, spinal cord, and disc herniation}\\
\multicolumn{6}{p{\linewidth}}{\textbf{Bold} values indicate the best performance for each metric.}\\
\end{tabular}
}
\end{table}

For diagnosis evaluation, we used the best segmentation results from each model across the five folds to compute clinical indicators automatically and compared them accordingly.

\begin{table*}[!ht]
\centering
\caption{Segmentation performance comparison.}
\label{table2}
\Large
\resizebox{2\columnwidth}{!}{
\begin{tabular}{lccccccccccccccccccccc}
\toprule
\multicolumn{22}{c}{\textbf{Renji dataset (internal test set)}}\\ 
\midrule
\multicolumn{2}{c}{\multirow{2}{*}{\textbf{Model}}} & \multicolumn{4}{c}{\textbf{IVD}} & \multicolumn{4}{c}{\textbf{V}} & \multicolumn{4}{c}{\textbf{SC}} & \multicolumn{4}{c}{\textbf{CSF}} & \multicolumn{4}{c}{\textbf{H}} \\
\cmidrule(lr){3-6} \cmidrule(lr){7-10} \cmidrule(lr){11-14} \cmidrule(lr){15-18} \cmidrule(lr){19-22}
&& \textbf{Dice↑} & \textbf{Jac↑} & \textbf{Prec↑} & \textbf{Rec↑} 
& \textbf{Dice↑} & \textbf{Jac↑} & \textbf{Prec↑} & \textbf{Rec↑} 
& \textbf{Dice↑} & \textbf{Jac↑} & \textbf{Prec↑} & \textbf{Rec↑} 
& \textbf{Dice↑} & \textbf{Jac↑} & \textbf{Prec↑} & \textbf{Rec↑} 
& \textbf{Dice↑} & \textbf{Jac↑} & \textbf{Prec↑} & \textbf{Rec↑} \\
\midrule
\multirow{6}{*}{\textbf{CNN}} 
& Standard Unet\cite{ronneberger2015u}  & 0.8956 & 0.8116 & 0.9031 & 0.8895 & 0.9305 & 0.8703 & \textbf{0.9234} & 0.9383 & 0.9038 & 0.8257 & 0.9052 & 0.9043 & 0.7793 & 0.6310 & 0.8166 & 0.7666 & 0.4165 & 0.2730 & 0.5002 & 0.3846 \\
& Attention Unet\cite{oktay2018attention} & 0.8957 & 0.8117 & 0.9037 & 0.8890 & 0.9306 & 0.8704 & 0.9224 & 0.9394 & 0.9030 & 0.8245 & \textbf{0.9072} & 0.9010 & 0.7780 & 0.6294 & 0.8219 & 0.7604 & 0.4168 & 0.2720 & 0.4893 & 0.3901 \\
& \textbf{Nested Unet\cite{zhou2018unet++}} & \textbf{0.8972} & \textbf{0.8142} & 0.9042 & \textbf{0.8920} & \textbf{0.9317} & \textbf{0.8723} & 0.9221 & \textbf{0.9421} & \textbf{0.9072} & \textbf{0.8313} & 0.9061 & 0.9107 & \textbf{0.7874} & \textbf{0.6414} & \textbf{0.8260} & 0.7745 & 0.4208 & 0.2763 & 0.4937 & 0.3952 \\
& CENet\cite{gu2019net} & 0.8792 & 0.7852 & 0.8849 & 0.8760 & 0.9201 & 0.8524 & 0.9120 & 0.9289 & 0.8771 & 0.7828 & 0.8743 & 0.8842 & 0.7345 & 0.5763 & 0.7347 & 0.7536 & 0.2815 & 0.1705 & 0.4618 & 0.2295 \\
& SegNet\cite{badrinarayanan2017segnet} & 0.8922 & 0.8064 & 0.8947 & 0.8917 & 0.9275 & 0.8652 & 0.9202 & 0.9360 & 0.8961 & 0.8131 & 0.8830 & \textbf{0.9120} & 0.7753 & 0.6253 & 0.7915 & \textbf{0.7807} & \textbf{0.4261} & \textbf{0.2802} & 0.4799 & \textbf{0.4146} \\
& Probabilistic Unet\cite{kohl2018probabilistic} & 0.8945 & 0.8100 & \textbf{0.9081} & 0.8832 & 0.9305 & 0.8703 & 0.9201 & 0.9417 & 0.8980 & 0.8161 & 0.9006 & 0.8981 & 0.7553 & 0.6029 & 0.8104 & 0.7322 & 0.3982 & 0.2588 & \textbf{0.5085} & 0.3577 \\
\midrule
\multirow{3}{*}{\textbf{Transformer}} 
& \textbf{TransUNet\cite{chen2021transunet}} & \textbf{0.8978} & \textbf{0.8150} & \textbf{0.8982} & \textbf{0.8985} & \textbf{0.9312} & \textbf{0.8715} & \textbf{0.9191} & \textbf{0.9442} & \textbf{0.8945} & \textbf{0.8107} & \textbf{0.9029} & 0.8889 & \textbf{0.7676} & \textbf{0.6170} & 0.7862 & \textbf{0.7714} & \textbf{0.4145} & \textbf{0.2711} & \textbf{0.4586} & \textbf{0.4102} \\
& SwinUNet\cite{cao2022swin} & 0.8747 & 0.7782 & 0.8707 & 0.8799 & 0.9155 & 0.8446 & 0.9077 & 0.9241 & 0.8883 & 0.8008 & 0.8865 & \textbf{0.8935} & 0.7639 & 0.6121 & 0.7888 & 0.7639 & 0.3042 & 0.1861 & 0.3846 & 0.2911 \\
& ViT\cite{dosovitskiy2020image} & 0.8672 & 0.7665 & 0.8748 & 0.8611 & 0.9174 & 0.8479 & 0.9119 & 0.9237 & 0.8879 & 0.8002 & 0.8958 & 0.8826 & 0.7567 & 0.6038 & \textbf{0.7933} & 0.7407 & 0.3237 & 0.2007 & 0.3799 & 0.3059 \\
\midrule
\multirow{1}{*}{\textbf{SAM}} 
& ViT-H SAM\cite{kirillov2023segany} & 0.8267 & 0.7093 & 0.9267 & 0.7510 & 0.9054 & 0.8291 & 0.8988 & 0.9139 & 0.3891 & 0.2560 & 0.5960 & 0.3223 & 0.5198 & 0.3576 & 0.3687 & 0.9069 & 0.1516 & 0.0903 & 0.3251 & 0.1253 \\
\midrule
\multirow{3}{*}{\textbf{nnUNet}} 
& nnUNet baseline\cite{isensee2021nnu} & 0.9423 & 0.8922 & 0.9514 & 0.9348 & 0.9596 & 0.9231 & 0.9516 & 0.9683 & 0.9261 & 0.8673 & 0.9266 & 0.9313 & 0.8302 & 0.7209 & 0.8610 & 0.8190 & 0.6411 & 0.4874 & 0.6742 & 0.6171 \\
& nnUNet+Edge Loss & 0.9481 & 0.9021 & 0.9585 & 0.9389 & 0.9635 & 0.9299 & 0.9533 & 0.9745 & 0.9436 & 0.8942 & \textcolor{red}{\textbf{0.9489}} & 0.9398 & 0.8324 & 0.7211 & 0.8698 & 0.8193 & 0.6469 & 0.4911 & 0.7087 & 0.6208 \\
& \textcolor{red}{\textbf{PG-nnUNet}} & \textcolor{red}{\textbf{0.9498}} & \textcolor{red}{\textbf{0.9052}} & \textcolor{red}{\textbf{0.9604}} & \textcolor{red}{\textbf{0.9404}} & \textcolor{red}{\textbf{0.9645}} & \textcolor{red}{\textbf{0.9317}} & \textcolor{red}{\textbf{0.9542}} & \textcolor{red}{\textbf{0.9755}} & \textcolor{red}{\textbf{0.9443}} & \textcolor{red}{\textbf{0.8955}} & 0.9483 & \textcolor{red}{\textbf{0.9419}} & \textcolor{red}{\textbf{0.8351}} & \textcolor{red}{\textbf{0.7250}} & \textcolor{red}{\textbf{0.8711}} & \textcolor{red}{\textbf{0.8229}} 
& \textcolor{red}{\textbf{0.6599}} & \textcolor{red}{\textbf{0.5103}} & \textcolor{red}{\textbf{0.7223}} & \textcolor{red}{\textbf{0.6337}} \\
\bottomrule
\multicolumn{22}{c}{\textbf{Ruijin dataset (external validation set)}}\\ 
\midrule
\multicolumn{2}{c}{\multirow{2}{*}{\textbf{Model}}} & \multicolumn{4}{c}{\textbf{IVD}} & \multicolumn{4}{c}{\textbf{V}} & \multicolumn{4}{c}{\textbf{SC}} & \multicolumn{4}{c}{\textbf{CSF}} & \multicolumn{4}{c}{\textbf{H}} \\
\cmidrule(lr){3-6} \cmidrule(lr){7-10} \cmidrule(lr){11-14} \cmidrule(lr){15-18} \cmidrule(lr){19-22}
&& \textbf{Dice↑} & \textbf{Jac↑} & \textbf{Prec↑} & \textbf{Rec↑} 
& \textbf{Dice↑} & \textbf{Jac↑} & \textbf{Prec↑} & \textbf{Rec↑} 
& \textbf{Dice↑} & \textbf{Jac↑} & \textbf{Prec↑} & \textbf{Rec↑} 
& \textbf{Dice↑} & \textbf{Jac↑} & \textbf{Prec↑} & \textbf{Rec↑} 
& \textbf{Dice↑} & \textbf{Jac↑} & \textbf{Prec↑} & \textbf{Rec↑} \\
\midrule
\multirow{6}{*}{\textbf{CNN}} 
& Standard Unet\cite{ronneberger2015u} & 0.9209 & 0.8539 & 0.9119 & 0.9308 & 0.9492 & 0.9039 & 0.9526 & 0.9463 & 0.9291 & 0.8684 & 0.9241 & 0.9355 & 0.8350 & 0.7211 & 0.8704 & 0.8070 & 0.5175 & 0.3600 & 0.5484 & 0.5156 \\
& Attention Unet\cite{oktay2018attention} & 0.9208 & 0.8538 & 0.9120 & 0.9305 & 0.9491 & 0.9037 & 0.9521 & 0.9465 & 0.9302 & 0.8702 & \textbf{0.9281} & 0.9335 & 0.8337 & 0.7194 & 0.8738 & 0.8021 & 0.5153 & 0.3584 & 0.5588 & 0.5050 \\
& \textbf{Nested Unet\cite{zhou2018unet++}} & \textbf{0.9215} & \textbf{0.8550} & 0.9105 & \textbf{0.9340} & \textbf{0.9494} & \textbf{0.9042} & 0.9494 & \textbf{0.9498} & \textbf{0.9311} & \textbf{0.8719} & 0.9265 & \textbf{0.9373} & \textbf{0.8438} & \textbf{0.7340} & 0.8743 & \textbf{0.8201} & \textbf{0.5270} & \textbf{0.3694} & 0.5538 & \textbf{0.5338} \\
& CENet\cite{gu2019net} & 0.9028 & 0.8235 & 0.8998 & 0.9073 & 0.9339 & 0.8767 & 0.9321 & 0.9364 & 0.9045 & 0.8271 & 0.9104 & 0.9010 & 0.7645 & 0.6252 & 0.7979 & 0.7404 & 0.2934 & 0.1825 & 0.5372 & 0.2200 \\
& SegNet\cite{badrinarayanan2017segnet} & 0.9141 & 0.8431 & 0.9129 & 0.9175 & 0.9424 & 0.8930 & 0.9450 & 0.9420 & 0.9175 & 0.8489 & 0.9167 & 0.9204 & 0.8199 & 0.6987 & 0.8275 & 0.8179 & 0.4816 & 0.3298 & 0.5451 & 0.4605 \\
& Probabilistic Unet\cite{kohl2018probabilistic} & 0.9198 & 0.8528 & \textbf{0.9235} & 0.9181 & 0.9472 & 0.9011 & \textbf{0.9544} & 0.9414 & 0.9230 & 0.8585 & 0.9263 & 0.9223 & 0.8091 & 0.6872 & \textbf{0.8863} & 0.7540 & 0.5103 & 0.3566 & \textbf{0.5769} & 0.4899 \\
\midrule
\multirow{3}{*}{\textbf{Transformer}}
& \textbf{TransUNet\cite{chen2021transunet}} & \textbf{0.9239} & \textbf{0.8591} & \textbf{0.9147} & \textbf{0.9337} & \textbf{0.9499} & \textbf{0.9051} & \textbf{0.9476} & \textbf{0.9525} & \textbf{0.9229} & \textbf{0.8580} & \textbf{0.9329} & 0.9146 & \textbf{0.8260} & \textbf{0.7086} & \textbf{0.8507} & 0.8074 & \textbf{0.5178} & \textbf{0.3615} & \textbf{0.5289} & \textbf{0.5412} \\
& SwinUNet\cite{cao2022swin} & 0.8967 & 0.8135 & 0.8897 & 0.9051 & 0.9327 & 0.8745 & 0.9327 & 0.9333 & 0.9201 & 0.8530 & 0.9164 & \textbf{0.9254} & 0.8214 & 0.7015 & 0.8198 & \textbf{0.8271} & 0.3432 & 0.2172 & 0.4102 & 0.3216 \\
& ViT\cite{dosovitskiy2020image} & 0.8903 & 0.8034 & 0.8788 & 0.9029 & 0.9334 & 0.8758 & 0.9348 & 0.9323 & 0.9116 & 0.8390 & 0.9196 & 0.9056 & 0.7989 & 0.6710 & 0.8297 & 0.7728 & 0.4204 & 0.2750 & 0.4190 & 0.4555 \\
\midrule
\multirow{1}{*}{\textbf{SAM}}
& ViT-H SAM\cite{kirillov2023segany} & 0.8308 & 0.7163 & 0.9439 & 0.7482 & 0.9001 & 0.8217 & 0.9043 & 0.9007 & 0.3456 & 0.2283 & 0.6264 & 0.2697 & 0.4830 & 0.3249 & 0.3394 & 0.8680 & 0.1082 & 0.0619 & 0.2444 & 0.0805 \\
\midrule
\multirow{3}{*}{\textbf{nnUNet}} 
& nnUNet baseline\cite{isensee2021nnu} & 0.9454 & 0.8970 & 0.9452 & 0.9457 & 0.9652 & 0.9332 & 0.9660 & 0.9646 & 0.9441 & 0.8951 & 0.9446 & 0.9445 & 0.8799 & 0.7894 & 0.8889 & 0.8724 & 0.6279 & 0.4739 & 0.6339 & 0.6356 \\
& nnUNet+Edge Loss & 0.9463 & 0.8987 & 0.9452 & \textcolor{red}{\textbf{0.9475}} & 0.9661 & 0.9348 & \textcolor{red}{\textbf{0.9678}} & 0.9645 & 0.9461 & 0.8987 & \textcolor{red}{\textbf{0.9503}} & 0.9428 & 0.8826 & 0.7939 & 0.8932 & 0.8734 & 0.6309 & 0.4799 & 0.6312 & 0.6477 \\
& \textcolor{red}{\textbf{PG-nnUNet}} & \textcolor{red}{\textbf{0.9472}} & \textcolor{red}{\textbf{0.9006}} & \textcolor{red}{\textbf{0.9480}} & 0.9466 & \textcolor{red}{\textbf{0.9666}} & \textcolor{red}{\textbf{0.9358}} & 0.9676 & \textcolor{red}{\textbf{0.9657}} & \textcolor{red}{\textbf{0.9463}} & \textcolor{red}{\textbf{0.8991}} & 0.9487 & \textcolor{red}{\textbf{0.9447}} & \textcolor{red}{\textbf{0.8848}} & \textcolor{red}{\textbf{0.7979}} & \textcolor{red}{\textbf{0.8952}} & \textcolor{red}{\textbf{0.8759}} & \textcolor{red}{\textbf{0.6431}} & \textcolor{red}{\textbf{0.4962}} & \textcolor{red}{\textbf{0.6451}} & \textcolor{red}{\textbf{0.6554}} \\
\bottomrule
\multicolumn{22}{l}{\textbf{Jac}=Jaccard; \textbf{Prec}=Precision; \textbf{Rec}=Recall}\\
\multicolumn{22}{l}{\textbf{IVD}=intervertebral disc; \textbf{V}=vertebral body; \textbf{CSF}=cerebral spinal fluid; \textbf{SC}=spinal cord; \textbf{H}=disc herniation}\\
\multicolumn{22}{l}{\textcolor{red}{\textbf{Bold}} values indicate the best performance among all models; \textbf{Bold} values indicate the best performance among CNN/Transformer models.}
\end{tabular}
}
\end{table*}

\begin{figure*}[!ht]
\centering
\includegraphics[width=1.8\columnwidth]{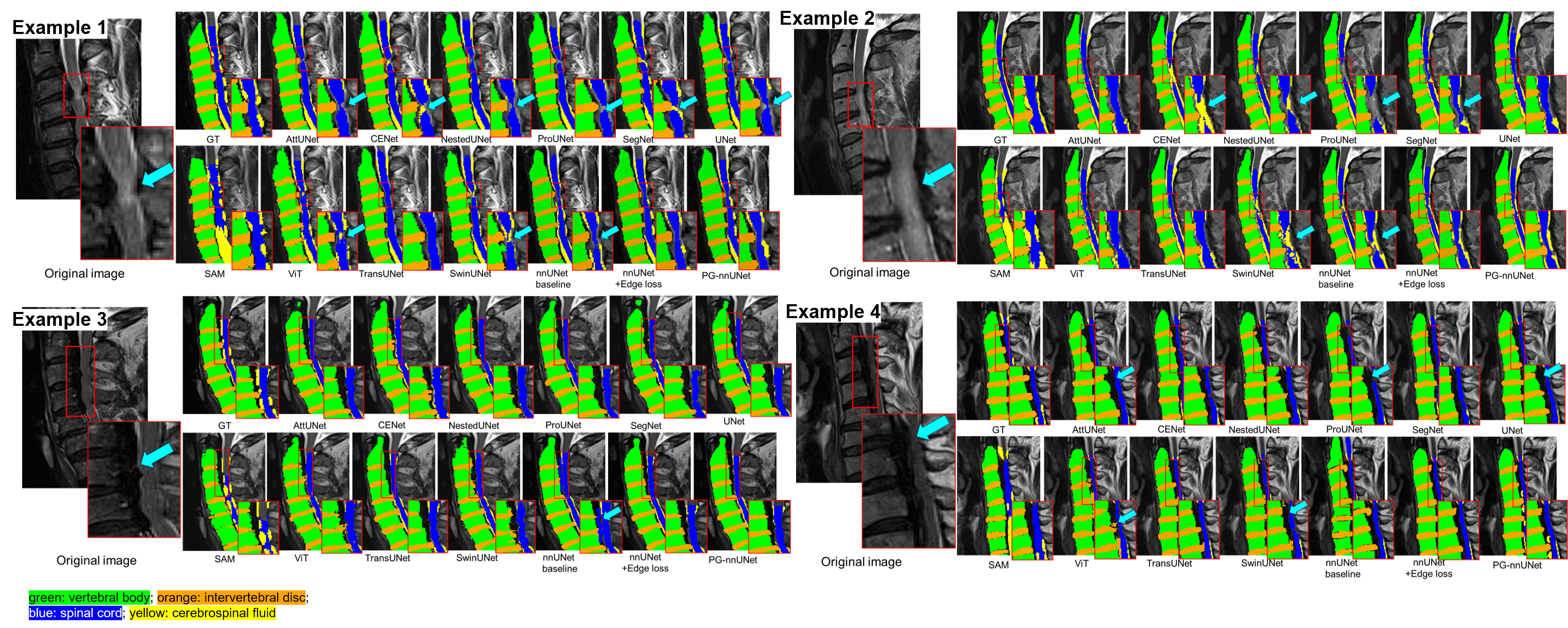}
\caption{\label{fig5} Examples of cervical spine segmentation results and comparisons across different models. The red boxes indicate pathological areas, such as T2 hyperintensity in the spinal cord and ossification of the posterior longitudinal ligament (OPLL).}
\end{figure*}

\subsection{Verification of expert-based diagnosis system}
A senior spine surgeon with ten years of experience and a highly experienced radiologist evaluated clinical indicators derived from the test set and external validation set. These indicators included C2-C7 Cobb Angle calculation, disc herniation localization, K-line status assessment (±), MSCC coefficient calculation, cervical spinal cord T2 hyperintensity detection, and Kang grading. Only segments above the C7 vertebral body were considered. After discussion to reach consensus, these indicators were compared to those from our  diagnosis system. Surgeons used ITK-SNAP\cite{py06nimg} for measurements, with evaluations matching the radiologist's report considered ground truth. System performance was assessed using Mean Absolute Error (MAE), Accuracy, Precision, Recall, and F1 score. For herniation localization, all five intervertebral discs (C2/3 to C6/7) across three layers were assessed, while other tasks focused on the middle layer due to potential absence or unclear visibility of spinal cord and cerebrospinal fluid in certain layers.

\subsection{Statistical Analysis}
Automated segmentation accuracy was evaluated by comparing results with manual annotations using metrics: Accuracy, Dice coefficient, Jaccard Index, Precision, and Recall. For diagnosis tasks, including herniation localization, K-line assessment, T2 hyperintensity detection, and Kang grading, we calculated Accuracy, Precision, Recall, and F1-score.

The Shapiro–Wilk test was employed to determine the normality of distributions. Continuous variables are expressed as the mean ± standard deviation (x ± s). The intraclass correlation coefficient (ICC) and Pearson correlation analysis were applied for comparing measurement data. To assess the diagnostic efficacy of the combined T2-MI and RSCI for cervical spinal T2 hyperintensity, we calculated the Receiver Operating Characteristic (ROC) Curve, Area Under the Curve (AUC), Sensitivity, Specificity, and Youden index. Statistical analysis was conducted using SPSS 25.0 software. The statistical significance threshold was set at \textit{P} \textless 0.05.

\section{Results}
\subsection{Spine Segmentation Model Performance}
\Cref{table2} presents the evaluation metrics for automatic segmentation models across five regions: intervertebral discs (IVD), vertebral bodies (V), cerebrospinal fluid (CSF), spinal cord (SC), and disc herniation (H). Note that H is derived from IVD using the method in \Cref{section:B}. Among CNN models, Nested UNet performed best, while TransUNet led among Transformer models. Therefore, Nested UNet and TransUnet were chosen for subsequent diagnosis comparisons.

nnUNet-based models outperformed CNN and Transformer models across all metrics for both the Renji internal test set and Ruijin external validation set. Among nnUNet variants, our PG-nnUNet achieved the highest Dice coefficients for IVD, V, SC, CSF, and H. For the internal test set, PG-nnUNet's Dice ranged from 0.8229 (CSF) to 0.9645 (V). For the external validation set, it ranged from 0.8848 (CSF) to 0.9666 (V). Despite H not being directly segmented by the network, PG-nnUNet demonstrated significant improvements, achieving the best performance and highest improvement ratio across all regions.

\begin{figure*}[!ht]
\centering
\includegraphics[width=1.8\columnwidth]{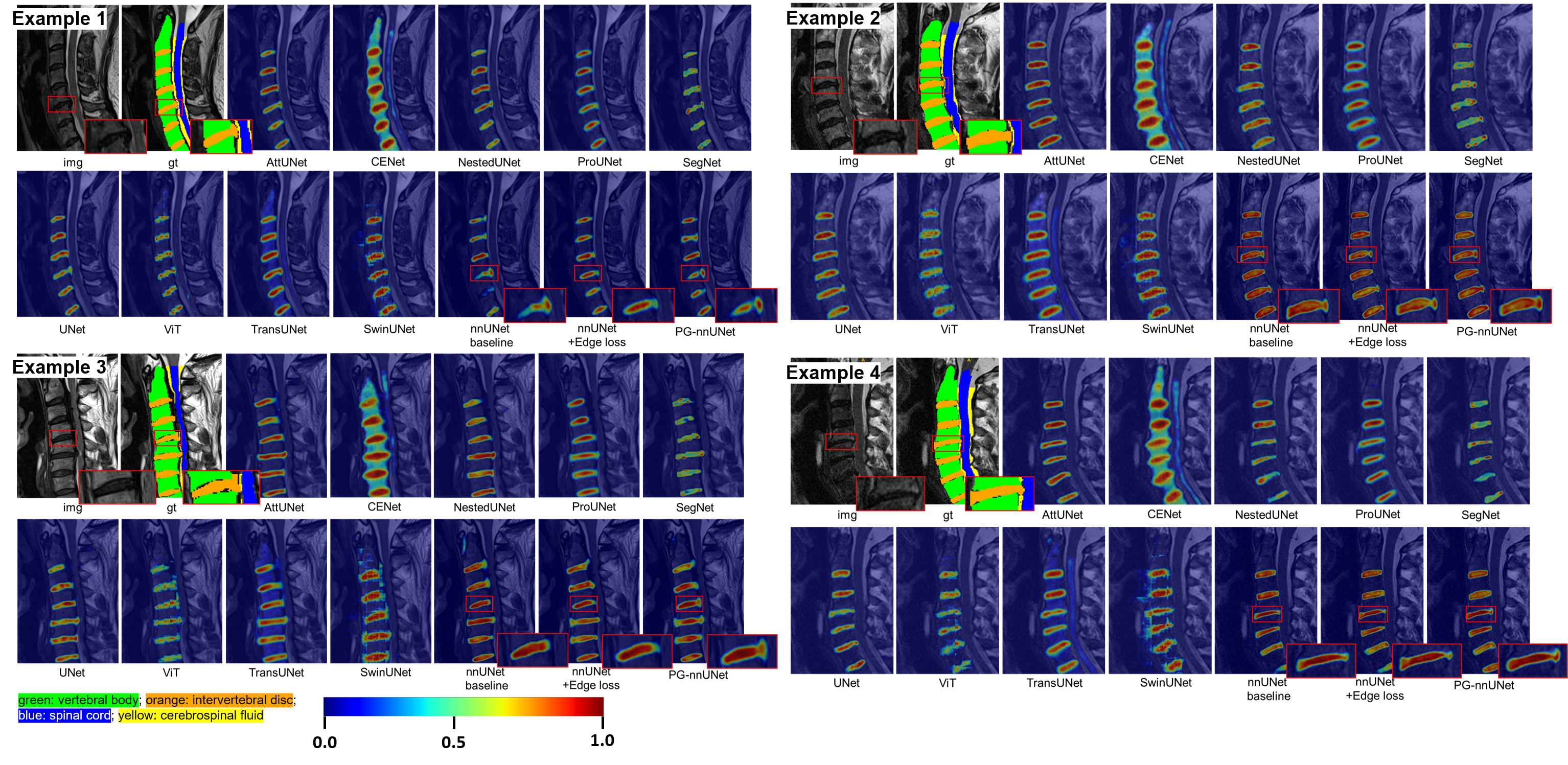}
\caption{\label{fig7} Visualization of the class activation mappings (CAMs) of intervertebral discs generated by Grad-CAM++\cite{chattopadhay2018grad}. The values are normalized to the range of 0-1. Regions with colors closer to 1 (hot colors) are more relevant to the segmentation results, while regions with colors closer to 0 (dark colors) are less relevant.}
\end{figure*}

\subsection{Visual Comparison of Segmentation Results} 
\Cref{fig5} visualizes segmentation predictions for several cases across different models. Example 1 and 2 highlight cases with T2 hyperintensity in the spinal cord (indicated by the red box), which is particularly challenging to predict accurately. Models like CENet, NestedUNet, and SwinUNet struggled to distinguish this region from CSF, while ProUNet misclassified it as background. In contrast, nnUNet with edge loss and our PG-nnUNet accurately segmented the region, preventing subsequent diagnostic errors. Examples 3 and 4 show ossification of the posterior longitudinal ligament (OPLL) cases. Many models inaccurately segmented the ossified region, with the nnUNet baseline even misclassifying it as part of the spinal cord. However, PG-nnUNet effectively detected these pathological regions.

We further analyzed model performance using Grad-CAM++\cite{chattopadhay2018grad} to visualize class activation mappings, focusing on intervertebral disc activation to assess attention to disc herniations. As shown in \Cref{fig7}, models like CENet failed to accurately activate target regions, while nnUNet-based models performed better. Compared to the nnUNet baseline, both edge loss-enhanced nnUNet and PG-nnUNet showed better attention to herniated areas, with PG-nnUNet notably highlighting herniated regions in Examples 1, 3, and 4. This improvement is attributed to the Pathology-Guided Detection Branch, which directs model focus toward pathological areas. Other models struggled to distinguish herniations from disc regions, leading to segmentation performance declines in these challenging areas.

\begin{table*}[!ht]
\centering
\caption{Comparison of diagnosis performance of Cobb angle and MSCC coefficient.}
\label{table3}
\resizebox{2\columnwidth}{!}{%
\begin{tabular}{cccccc}
\toprule
\multirow{2}{*}{\textbf{Metric}} & \multirow{2}{*}{\textbf{Model}} & \multicolumn{2}{c}{\textbf{Renji dataset (internal test set)}} & \multicolumn{2}{c}{\textbf{Ruijin dataset (external validation set)}} \\
\cmidrule(lr){3-4} \cmidrule(lr){5-6}
\multicolumn{2}{c}{} & \textbf{MAE↓} & \textbf{ICC↑} & \textbf{MAE↓} & \textbf{ICC↑} \\
\midrule
\multirow{5}{*}{\textbf{C2-C7 Cobb Angle(°)}} & Nested Unet\cite{zhou2018unet++} & 3.88±4.42 & 0.801(0.722-0.859) ** & 4.43±5.03 & 0.678(0.517-0.786) **\\
& TransUNet\cite{chen2021transunet} & 3.30±2.66 & 0.887(0.840-0.922) ** & 4.26±5.84 & 0.780(0.657-0.857) ** \\
& nnUNet baseline\cite{isensee2021nnu} & 2.59±2.10 & 0.931(0.901-0.952) ** & 3.29±4.32 & 0.842(0.783-0.895) **\\
& nnUNet+Edge loss & 2.64±2.15 & 0.926(0.893-0.948) ** & 2.89±2.24 & \textbf{0.923(0.884-0.949)} **\\
& \textbf{PG-nnUNet}& \textbf{2.44±1.93} & \textbf{0.940(0.914-0.959)} ** & \textbf{2.84±2.32} & 0.922(0.883-0.948) ** \\
\midrule
\multirow{5}{*}{\textbf{MSCC Coefficient(\%)}} & Nested Unet\cite{zhou2018unet++} & 10.51±7.87 & 0.666(0.545-0.760) *  & 8.90±8.25 & 0.487(0.236-0.656) *\\
& TransUNet\cite{chen2021transunet} & 8.32±8.62 & 0.712(0.603-0.795) ** & 7.68±6.65 & 0.556(0.337-0.703) ** \\
& nnUNet baseline\cite{isensee2021nnu} & 5.47±5.28 & 0.876(0.823-0.914) ** & 6.05±4.96 & 0.711(0.566-0.808) ** \\
& nnUNet+Edge loss & 4.90±4.07 & 0.919(0.883-0.944) ** & 5.36±4.92 & 0.801(0.701-0.868) **\\
& \textbf{PG-nnUNet} & \textbf{3.60±2.37} & \textbf{0.961(0.943-0.973)} ** & \textbf{3.93±3.29} & \textbf{0.878(0.817-0.919)} ** \\
\bottomrule
\multicolumn{6}{l}{\textbf{MAE}: Mean Absolute Error (Mean±SD); \textbf{ICC}: Interclass Correlation Coefficient (95\% CI); SD: Standard Deviation; CI: Confidence Interval}\\
\multicolumn{4}{l}{*:  \textit{P}  \textless 0.001; **:  \textit{P}  \textless 0.0001} \\
\multicolumn{4}{l}{\textbf{Bold} values indicate the best performance for each metric.}
\end{tabular}
}
\end{table*}

\begin{figure}[!ht]
\centering
\includegraphics[width=\columnwidth]{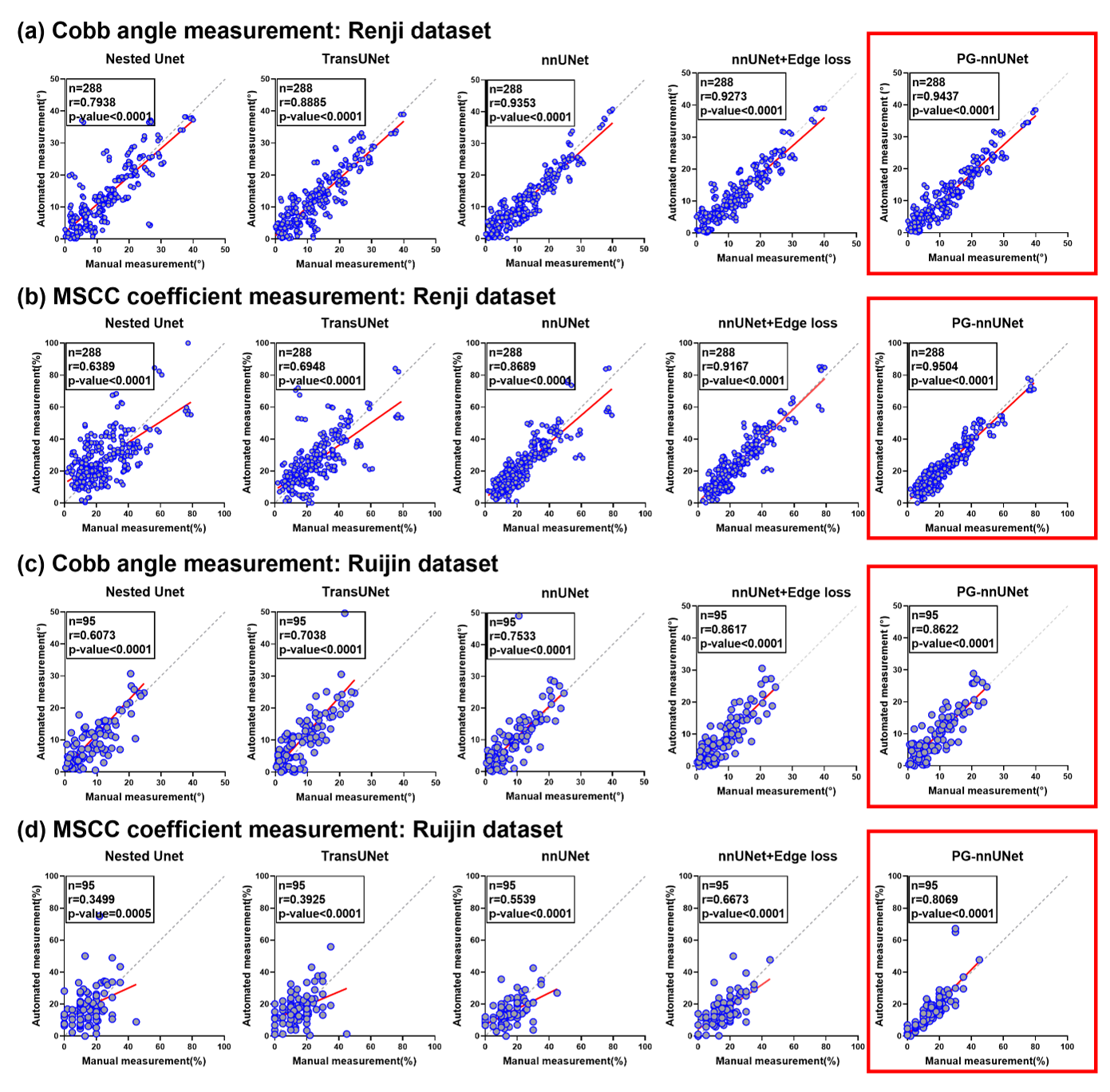}
\caption{\label{fig6} Correlations between manual and automated measurements for Cobb angle and MSCC coefficients. Five segmentation models were evaluated to assess the impact of segmentation on subsequent diagnosis. Pearson correlation coefficients (r) were used to quantify the strength of the correlation.}
\end{figure}

\subsection{Accuracy of Cobb angle and MSCC coefficient calculation}
As shown in \Cref{table3}, nnUNet-based models achieved superior diagnosis performance with lower Mean Absolute Error (MAE) and higher Intraclass Correlation Coefficients (ICC) for both Cobb angle and MSCC coefficient calculations compared to Nested UNet and TransUNet. For the internal test set, PG-nnUNet reduced the MAE for Cobb angle from $2.59^{\circ}$ to $2.44^{\circ}$, and for MSCC from 5.47\% to 3.60\%. For the external validation set, PG-nnUNet achieved the lowest MAE of $2.84^{\circ}$ for Cobb angle and 3.93\% for MSCC. All models showed high consistency with expert diagnoses, with most \textit{p}-values \textless 0.0001 (except Nested UNet, \textit{p} \textless 0.001). PG-nnUNet obtained the highest ICC coefficients for both metrics in the internal test set. As shown in \Cref{fig6}, PG-nnUNet demonstrated the strongest agreement between manual and automated measurements, achieving the highest Pearson correlation coefficients for Cobb angle (Renji: 0.9437, Ruijin: 0.8622) and MSCC (Renji: 0.9504, Ruijin: 0.8069).

\begin{table*}[!ht]
\centering
\caption{Comparison of diagnosis performance of Herniation Localization, K-line Assessment (±), and T2 Hyperintensity Detection.}
\label{table4}
\resizebox{2\columnwidth}{!}{%
\begin{tabular}{llcccccccc}
\hline
\multirow{2}{*}{\textbf{Task}} & \multirow{2}{*}{\textbf{Model}}  & \multicolumn{4}{c}{\textbf{Renji dataset (internal test set)}} & \multicolumn{4}{c}{\textbf{Ruijin dataset (external validation set)}} \\
\cmidrule(lr){3-6} \cmidrule(lr){7-10}       
\multicolumn{2}{c}{} & \textbf{Acc↑} & \textbf{Prec↑} & \textbf{Rec↑} & \textbf{F1↑} & \textbf{Acc↑} & \textbf{Prec↑} & \textbf{Rec↑} & \textbf{F1↑} \\ \hline
\multirow{5}{*}{\textbf{Herniation Localization}} 
& Nested Unet & 0.7402 & \textbf{0.8863} & 0.7162 & 0.7922 & 0.8288 & \textbf{0.8897} & 0.8649 & 0.8771 \\ 
& TransUNet & 0.7668 & 0.8770 & 0.7709 & 0.8206 & 0.8182 & 0.8425 & 0.9136 & 0.8766 \\ 
& nnUNet baseline & 0.8056 & 0.8201 & 0.9209 & 0.8676 & 0.8344 & 0.8521 & 0.9265 & 0.8877 \\ 
& nnUNet+Edge loss & 0.8051 & 0.8173 & \textbf{0.9250} & 0.8678 & \textbf{0.8547} & 0.8839 & 0.9146 & 0.8990 \\ 
& \textbf{PG-nnUNet} & \textbf{0.8075} & 0.8209 & 0.9230 &\textbf{0.8690}  & 0.8526 & 0.8633 & \textbf{0.9404} &\textbf{0.9002}   \\ \hline
\multirow{5}{*}{\textbf{K-line Assessment (±)}} 
& Nested Unet & 0.9375 & 0.8462 & 0.6111 & 0.7097 & - & - & - & - \\ 
& TransUNet & 0.9375 & 0.8462 & 0.6111 & 0.7097 & - & - & - & - \\ 
& nnUNet baseline & 0.9444 & 0.9167 & 0.6111 & 0.7333 & - & - & - & - \\ 
& nnUNet+Edge loss & 0.9583 & \textbf{1.0000} & 0.6667 & 0.8000 & - & - & - & - \\ 
& \textbf{PG-nnUNet} & \textbf{0.9722} & \textbf{1.0000} & \textbf{0.7778} & \textbf{0.8750} & - & - & - & - \\ \hline
\multirow{5}{*}{\textbf{T2 Hyperintensity Detection}} 
& Nested Unet & 0.7847 & 0.7959 & 0.8764 & 0.8342 & 0.8955 & 0.0526 & 0.1667 & 0.0800 \\ 
& TransUNet & 0.7847 & 0.7900 & 0.8876 & 0.8360 & 0.9273 & 0.1429 & 0.3333 & 0.2000 \\ 
& nnUNet baseline & 0.8264 & 0.8200 & 0.9213 & 0.8677 & 0.9273 & 0.1739 & 0.4444 & 0.2500 \\ 
& nnUNet+Edge loss & 0.8542 & 0.8400 & 0.9438 & 0.8889 & 0.9364 & 0.2143 & 0.5000 & 0.3000 \\ 
& \textbf{PG-nnUNet} & \textbf{0.8611} & \textbf{0.8416} & \textbf{0.9551} & \textbf{0.8947} & \textbf{0.9379} & \textbf{0.2553} & \textbf{0.6667} & \textbf{0.3692} \\ \hline
\multicolumn{10}{l}{\textbf{Acc}=Accuracy; \textbf{Prec}=Precision; \textbf{Rec}=Recall; \textbf{F1}=F1-score}\\
\multicolumn{4}{l}{\textbf{Bold} values indicate the best performance for each metric.}\\
\end{tabular}
}
\end{table*}

\begin{figure}[!ht]
\centering
\includegraphics[width=\columnwidth]{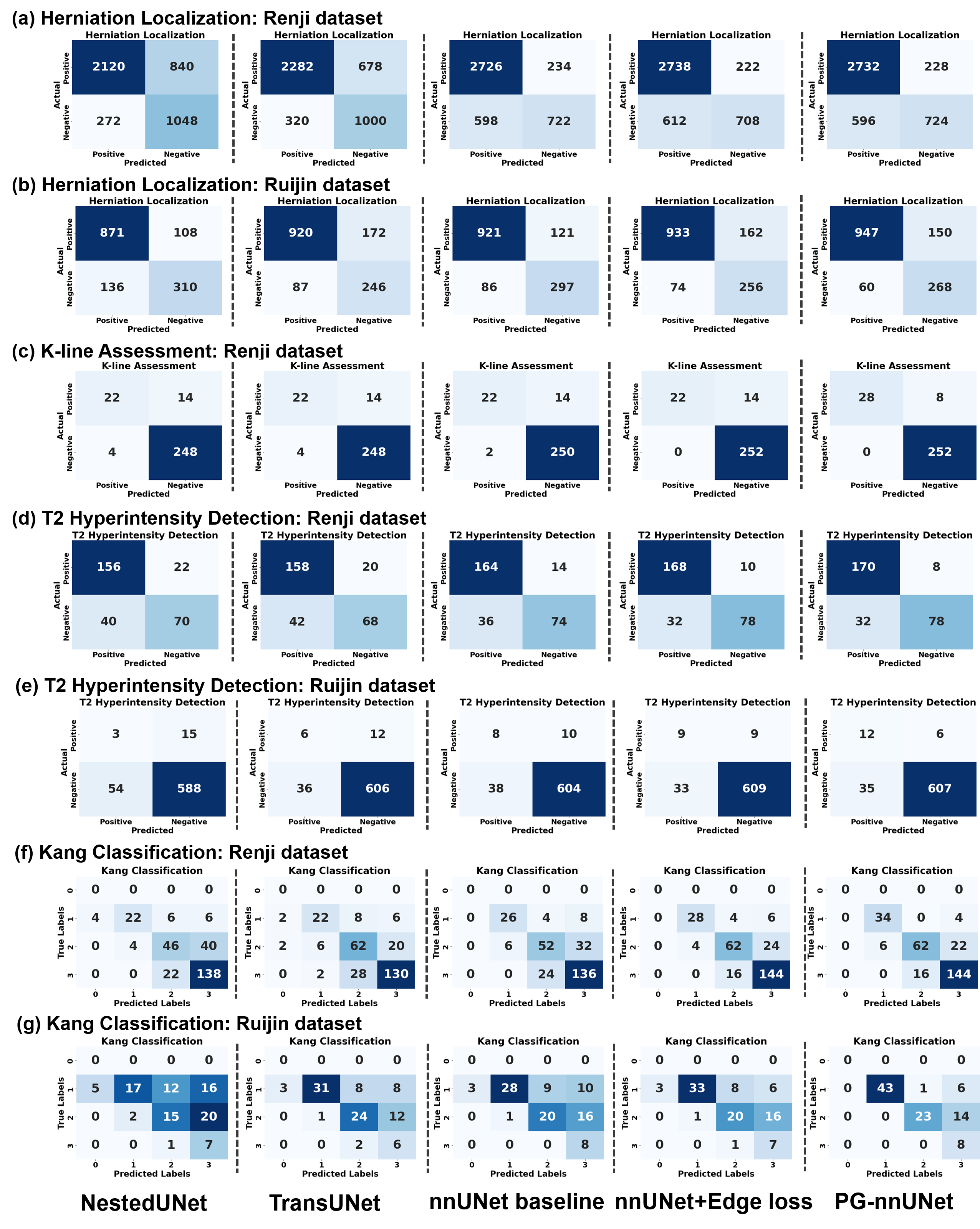}
\caption{\label{fig8} Confusion matrices on different models for herniation localization, K-line assessment, T2 hyperintensity detection, and Kang grading.}
\end{figure}

\begin{figure}[!ht]
\centering
\includegraphics[width=\columnwidth]{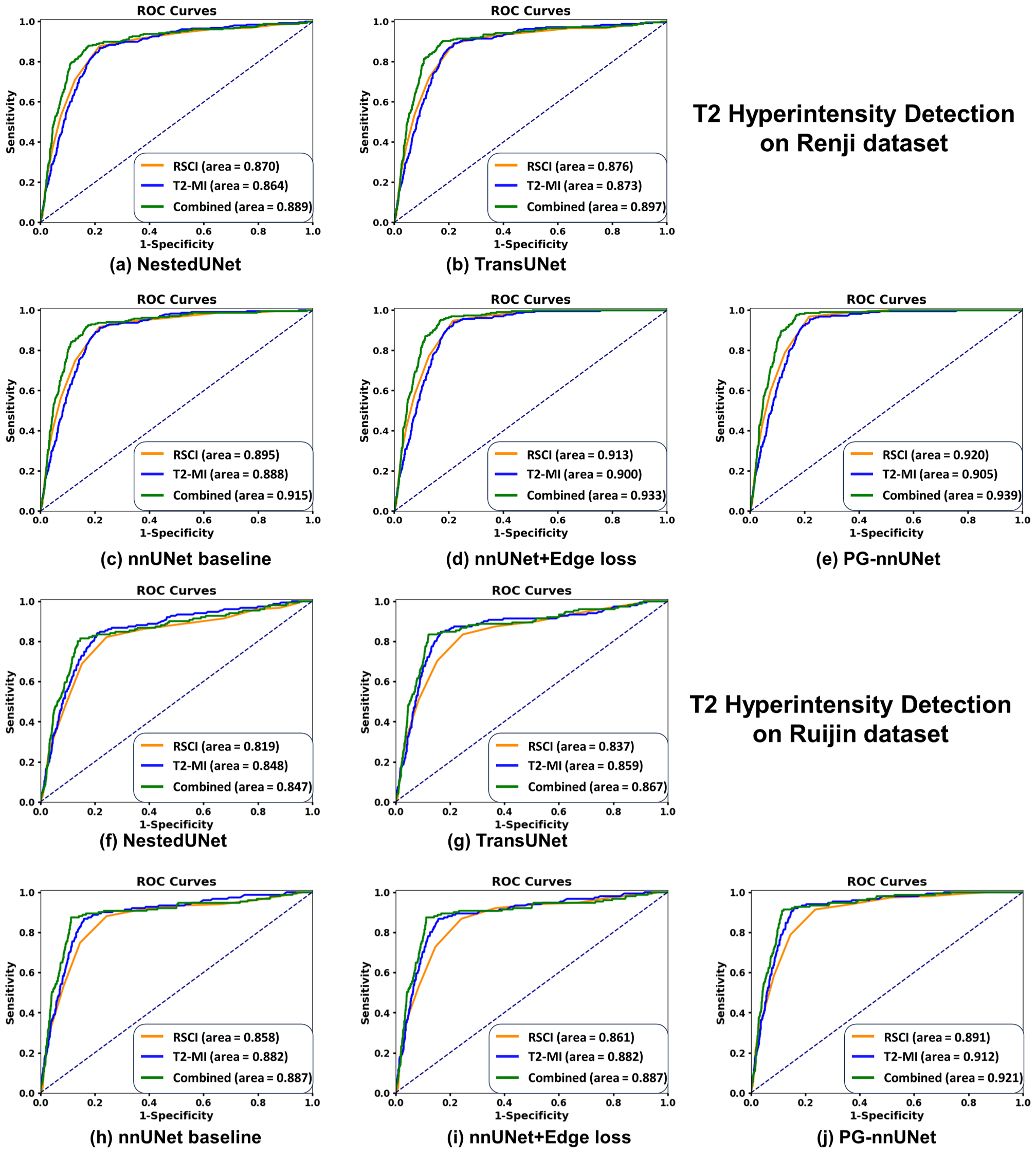}
\caption{\label{fig9}ROC curves for T2-MI alone, RSCI alone, and the combined indicators, demonstrating the comparative diagnostic performance for T2 hyperintensity in cervical myelopathy lesions. (a-e): T2 hyperintensity detection performance on Renji dataset; (f-j): T2 hyperintensity detection performance on Ruijin dataset.}
\end{figure}

\subsection{Evaluation of Other Diagnostic Indicators}
The diagnostic accuracy, precision, recall, and F1 score for automated herniation localization, K-line assessment (±), and T2 hyperintensity detection across different models are presented in \Cref{table4}.
In the Herniation Localization task, PG-nnUNet achieved the highest F1 score of 0.8690 and 0.9002 across the two datasets, balancing precision and recall effectively. While Nested UNet showed the highest precision, its lower recall indicated reduced sensitivity. In the K-line Assessment (±) task (using only the Renji dataset due to the absence of K-line (+) cases in the Ruijin dataset), PG-nnUNet outperformed other models with an accuracy of 0.9722 and an F1 score of 0.8750, supported by perfect precision and a high recall of 0.7778. Although nnUNet+Edge loss also performed well, its F1 score remained lower than PG-nnUNet's. Other models like Nested UNet and TransUNet showed lower recall and F1 scores. In the T2 Hyperintensity Detection task, PG-nnUNet achieved the highest scores across all metrics. The combination of T2-MI and RSCI contributed to its superior performance, with recall exceeding 0.95 and F1 score nearing 0.90. Performance for all models declined in the external validation set due to domain bias. Confusion matrices for the five models are shown in \Cref{fig8}.

To further evaluate T2 hyperintensity detection, Receiver Operating Characteristic (ROC) curves were plotted for T2-MI, RSCI, and their combination in \Cref{fig9}. The combination of both indices achieved the best results with improved Area Under the Curve (AUC). PG-nnUNet recorded the highest AUC values (Renji: 0.939, Ruijin: 0.921) when both indices were combined. The optimal diagnostic cutoff values were determined as $RSCI \geq 1.2$ and $T2-MI \geq 23.7$ based on Renji dataset, with a Youden index of 0.807.

\subsection{Performance of Automatic Kang Grading}
\label{sec:kang_results}
Table \ref{table_grading} details the performance of automatic Kang grading across four stenosis levels. Our PG-nnUNet achieved best results in both internal (Renji) and external (Ruijin) test sets. In Renji dataset, all nnUNet-based models achieved 100\% precision/recall (no false negatives/positives) for Kang 0. PG-nnUNet attained the highest recall (0.8947) and F1-score (0.8718) for Kang 1, reducing missed diagnoses compared to Nested UNet (recall=0.5789). PG-nnUNet also achieved great improvement in Kand 2 and 3 with highest F1-scores 0.7317 and 0.8727. In the Ruijin dataset, only PG-nnUNet perfectly identified healthy cases for Kang 0. Additionally, PG-nnUNet achieved the highest values in all F1-scores and most metrics for Kang 1-3, demonstrating its robustness and excellent performance in the external validation set. 100\% sensitivity in the Kang 3 grading ensured no severe cases were missed, despite precision challenges from dataset bias. Notably, Kang 2 remained challenging across models in both datasets ($F1 \leq 0.7541$), likely due to ambiguous cord deformation patterns. This aligns with clinical observations where Grade 1/2 distinctions often require axial MRI confirmation.

\begin{table*}[!ht]
\centering
\caption{Kang grading performance comparison.}
\label{table_grading}
\Large
\resizebox{2\columnwidth}{!}{
\begin{tabular}{lccccccccccccccccc}
\toprule
\multicolumn{17}{c}{\textbf{Renji dataset (internal test set)}}\\ 
\midrule
\multicolumn{1}{c}{\multirow{2}{*}{\textbf{Model}}} & 
\multicolumn{3}{c}{\textbf{Kang 0}} & 
\multicolumn{3}{c}{\textbf{Kang 1}} & 
\multicolumn{3}{c}{\textbf{Kang 2}} & 
\multicolumn{3}{c}{\textbf{Kang 3}} & 
\multicolumn{4}{c}{\textbf{Macro average}} \\
\cmidrule(lr){2-4} \cmidrule(lr){5-7} \cmidrule(lr){8-10} \cmidrule(lr){11-13} \cmidrule(lr){14-17}
& \textbf{Prec↑} & \textbf{Rec↑} & \textbf{F1↑} & \textbf{Prec↑} & \textbf{Rec↑} & \textbf{F1↑} & \textbf{Prec↑} & \textbf{Rec↑} & \textbf{F1↑} & \textbf{Prec↑} & \textbf{Rec↑} & \textbf{F1↑} & \textbf{Acc↑} & \textbf{Prec↑} & \textbf{Rec↑} & \textbf{F1↑}\\
\midrule
Nested Unet & 0.0000 & 0.0000 & 0.0000 & 0.8462 & 0.5789 & 0.6875 & 0.6111 & 0.5000 & 0.5500 & 0.7500 & 0.8625 & 0.8023 & 0.7133 & 0.5518 & 0.4854 & 0.5100 \\
TransUNet & 0.0000 & 0.0000 & 0.0000 & 0.7333 & 0.5789 & 0.6471 & 0.6250 & \textbf{0.6818} & 0.6522 & 0.8333 & 0.8125 & 0.8228 & 0.7413 & 0.5479 & 0.5183 & 0.5305 \\
nnUNet baseline & \textbf{1.0000} & \textbf{1.0000} & \textbf{1.0000} & 0.8125 & 0.6842 & 0.7429 & 0.6410 & 0.5682 & 0.6024 & 0.7727 & 0.8500 & 0.8095 & 0.7413 & 0.7421 & 0.7008 & 0.7183 \\
nnUNet+Edge loss & \textbf{1.0000} & \textbf{1.0000} & \textbf{1.0000} & \textbf{0.8750} & 0.7368 & 0.8000 & 0.7500 & \textbf{0.6818} & 0.7143 & 0.8276 & \textbf{0.9000} & 0.8623 & 0.8112 & 0.8175 & 0.7729 & 0.7922 \\
\textbf{PG-nnUNet} & \textbf{1.0000} & \textbf{1.0000} & \textbf{1.0000} & 0.8500 & \textbf{0.8947} & \textbf{0.8718} & \textbf{0.7895} & \textbf{0.6818} & \textbf{0.7317} & \textbf{0.8471} & \textbf{0.9000} & \textbf{0.8727} & \textbf{0.8322} & \textbf{0.8288} & \textbf{0.8255} & \textbf{0.8254} \\
\midrule
\multicolumn{17}{c}{\textbf{Ruijin dataset (external validation set)}}\\ 
\midrule
\multicolumn{1}{c}{\multirow{2}{*}{\textbf{Model}}} & 
\multicolumn{3}{c}{\textbf{Kang 0}} & 
\multicolumn{3}{c}{\textbf{Kang 1}} & 
\multicolumn{3}{c}{\textbf{Kang 2}} & 
\multicolumn{3}{c}{\textbf{Kang 3}} & 
\multicolumn{4}{c}{\textbf{Macro average}} \\
\cmidrule(lr){2-4} \cmidrule(lr){5-7} \cmidrule(lr){8-10} \cmidrule(lr){11-13} \cmidrule(lr){14-17}
& \textbf{Prec↑} & \textbf{Rec↑} & \textbf{F1↑} & \textbf{Prec↑} & \textbf{Rec↑} & \textbf{F1↑} & \textbf{Prec↑} & \textbf{Rec↑} & \textbf{F1↑} & \textbf{Prec↑} & \textbf{Rec↑} & \textbf{F1↑} & \textbf{Acc↑} & \textbf{Prec↑} & \textbf{Rec↑} & \textbf{F1↑}\\
\midrule
Nested Unet & 0.0000 & 0.0000 & 0.0000 & 0.8947 & 0.3400 & 0.4928 & 0.5357 & 0.4054 & 0.4615 & 0.1628 & 0.8750 & 0.2745 & 0.4105 & 0.3983 & 0.4051 & 0.3072 \\
TransUNet & 0.0000 & 0.0000 & 0.0000 & 0.9688 & 0.6200 & 0.7561 & 0.7059 & \textbf{0.6486} & 0.6761 & 0.2308 & 0.7500 & 0.3529 & 0.6421 & 0.4764 & 0.5047 & 0.4463 \\
nnUNet baseline & 0.0000 & 0.0000 & 0.0000 & 0.9655 & 0.5600 & 0.7089 & 0.6897 & 0.5405 & 0.6061 & 0.2353 & \textbf{1.0000} & 0.3810 & 0.5895 & 0.4726 & 0.5251 & 0.4240 \\
nnUNet+Edge loss & 0.0000 & 0.0000 & 0.0000 & 0.9706 & 0.6600 & 0.7857 & 0.6897 & 0.5405 & 0.6061 & 0.2414 & 0.8750 & 0.3784 & 0.6316 & 0.4754 & 0.5189 & 0.4425 \\
\textbf{PG-nnUNet} & \textbf{1.0000} & \textbf{1.0000} & \textbf{1.0000} & \textbf{1.0000} & \textbf{0.8600} & \textbf{0.9247} & \textbf{0.9583} & 0.6216 & \textbf{0.7541} & \textbf{0.2857} & \textbf{1.0000} & \textbf{0.4444} & \textbf{0.7789} & \textbf{0.7480} & \textbf{0.8272} & \textbf{0.7078} \\
\bottomrule
\multicolumn{17}{l}{\textbf{Kang 0}: no stenosis; \textbf{Kang 1}: mild stenosis; \textbf{Kang 2}: moderate stenosis; \textbf{Kang 3}: severe stenosis.}\\
\multicolumn{17}{l}{\textbf{Prec}=Precision; \textbf{Rec}=Recall; \textbf{F1}=F1-score; \textbf{Acc}=Accuracy}\\
\multicolumn{17}{l}{\textbf{Bold} values indicate the best performance among all models.}
\end{tabular}
}
\end{table*}

\begin{table*}[!ht]
\centering
\caption{Summary of diagnostic indicators, methods, and advantages of our model.}
\label{table_diagnosis}
\renewcommand{\arraystretch}{1.5} 
\resizebox{\textwidth}{!}{ 
\begin{tabular}{|>{\centering\arraybackslash}m{4cm}|>{\centering\arraybackslash}m{5cm}|>{\centering\arraybackslash}m{4cm}|>{\centering\arraybackslash}m{4.5cm}|}
\hline
\textbf{Task} & \textbf{Significance} & \textbf{Method/Indicator} & \textbf{Model Advantages} \\
\hline
\textbf{Cobb Angle Calculation} & 
Key indicator for evaluating scoliosis and kyphosis. &
Calculated by extending lines along C2 and C7 endplates.&
\multirow{5}{4.5cm}{Manual measurement is time-consuming and prone to errors, especially for less experienced physicians. Our expert-based automated approach enhances efficiency, provides more accurate results, and offers quantifiable indicators with visualized outputs, improving interpretability.} \\
\cline{1-3}
\textbf{MSCC Calculation} & 
Quantifies spinal cord compression from disc herniation.&
\( MSCC = 1 - \frac{D_i}{\left(\frac{D_a + D_b}{2}\right)} \). & \\ 
\cline{1-3}
\textbf{K-line Plotting and Assessment} & 
Guides surgical planning by showing spinal canal-herniation relationship.&
Automatically draws the K-line connecting C2 and C7. & \\ 
\cline{1-3}
\textbf{Herniation Localization} & 
Identifies disc protrusion for better treatment planning. &
Segmenting and splitting herniated discs. & \\ 
\cline{1-3}
\textbf{T2 Hyperintensity} & 
Detects spinal cord pathologies like edema or degeneration. &
T2-MI and RSCI (Relative Signal Change Index). & \\ 
\cline {1-3}
\textbf {Kang Grading} &
Categorizes cervical spinal stenosis severity into four grades (0-3). &
Rule-based assessment using our calculated quantitative metrics. & \\
\hline
\end{tabular}
}
\end{table*}

\section{Discussion}
Our PG-nnUNet demonstrated high performance in segmenting four key cervical spinal anatomies, achieving Dice values over 0.90 in intervertebral discs, vertebral bodies (V), and spinal cord (SC), meeting clinical precision requirements. This accuracy is likely due to the regular structures and high contrast of these anatomies in T2-weighted MR images. However, cerebrospinal fluid (CSF) segmentation performance was slightly lower, possibly because of its irregular shape and limited grayscale contrast with adjacent tissues, which can be complicated by disc herniations displacing or obscuring the CSF.

PG-nnUNet outperformed traditional CNN and Transformer models across all evaluation metrics for the four anatomical regions. By incorporating edge loss and a Pathology-Guided Detection Branch, PG-nnUNet significantly improved upon the original nnUNet. Notably, disc herniation segmentation, crucial for accurate diagnosis, saw the greatest improvement, with Dice scores increasing from 0.6411 to 0.6599 (Renji) and 0.6279 to 0.6431 (Ruijin). These enhancements directly improved the accuracy of subsequent diagnostic processes.

While the original nnUNet excels in many medical segmentation tasks through its self-configuration strategy, it falls short in focusing on pathology-specific areas like disc herniations. Errors in these small, complex regions can lead to diagnostic inaccuracies. To address this, we developed a method leveraging expert knowledge to generate pathology heatmaps without additional manual annotation. Unlike traditional heatmaps, ours use a modified Gaussian distribution to represent pathological impact pixel-wise, guiding the network to focus on critical yet difficult-to-segment regions. Our PG-nnUNet's dual-branch architecture and multi-task strategy balance segmentation accuracy and pathology focus while retaining the original architecture's self-configuration capabilities. The pathology heatmap prediction acts as an auxiliary task that provides spatial guidance, helping the segmentation branch focus on diagnostically critical and often challenging pathological regions, leading to the observed improvements.

Our ultimate goal is to construct a precise and fully automatic diagnosis system for cervical spondylosis, requiring three critical components: 1) accurate segmentation of relevant cervical spinal anatomies, especially pathological regions; 2) expert guidance in selecting optimal diagnostic indicators, configuring diagnostic parameters, and proposing novel indicators; and 3) accurate localization, plotting, and calculation of diagnostic indicators, ultimately generating a comprehensive report. After that, complete evaluation and comparison of different methods with experts measurement are also crucial to ensure the reliability and accuracy of the whole system. Combined with clinical knowledge and expert guidance, we have constructed a detailed diagnosis framework including several diagnosis tasks and key indicators, as described in \Cref{section:D}.

For Cobb angle and MSCC evaluation, nnUNet-based models consistently outperformed CNN-based and Transformer-based models. PG-nnUNet exhibited even greater improvements in diagnostic accuracy than in segmentation. The average MAE for Cobb angle calculation using PG-nnUNet was $2.32^\circ \pm 2.04^\circ$, within the clinically acceptable error range of $3^\circ$ to $10^\circ$\cite{sardjono2013automatic}. For MSCC, the MAE decreased from 5.47\%±5.28\% to 3.60\%±2.37\% (34\%↑), and the ICC coefficient increased from 0.876 to 0.961 (9.7\%↑). These enhancements are attributed to the precise segmentation of pathological areas with the integration of pathology-guided prior knowledge.

In the Herniation Localization task, PG-nnUNet demonstrated the most balanced performance with the highest F1 score (0.8690) and the best overall accuracy (0.8075), effectively balancing precision and recall. This makes it well-suited for clinical use where both detecting herniations and avoiding false positives are important. Nested Unet achieved the highest precision (0.8863) but had a low recall (0.7162), limiting its sensitivity and potentially leading to missed diagnoses. In the K-line Assessment (±) task, our PG-nnUNet-based system achieved the highest accuracy (0.9722) and F1 score (0.8750), with perfect precision (1.0000) and a high recall (0.7778), precisely identifying cases where the K-line intersects with depressor structures. The perfect precision indicates that the model effectively avoids false positives, which is crucial in clinical scenarios where misidentifying K-line intersections could lead to inappropriate surgical interventions.

For T2 Hyperintensity detection, PG-nnUNet outperformed all other models with the highest accuracy (0.8611) and F1 score (0.8947). Combining T2-MI and RSCI contributed to its superior performance, effectively identifying abnormal signal intensity patterns. This is critical in clinical settings where early detection of spinal cord pathology can significantly impact treatment outcomes. The high recall (0.9551) ensures most pathological hyperintensity cases are detected, reducing missed diagnoses.

The Kang grading system demonstrates established clinical relevance, with severity levels positively correlating to myelopathy-related neurologic deterioration and surgical indication rates\cite{kang2011new}. Our automated implementation provides standardized qualitative severity assessment through discrete stenosis grading, reducing inter-rater variability while enhancing detection of mild-to-moderate cases. Despite inherent limitations from image quality and anatomical variations, integrating Kang grading with our model's quantitative parameters (e.g., MSCC, T2 hyperintensity) enables synergistic comprehensive evaluation, which optimizes holistic disease quantification and patient management.

A summary of the diagnostic indicators, methods, and advantages of our model is provided in \Cref{table_diagnosis}.

\section{Conclusion and Limitations}
We introduced an AI-assisted Expert-based Diagnosis System for cervical spondylosis, featuring the novel PG-nnUNet architecture. By integrating edge loss and a Pathology-Guided Detection Branch, our model achieved high segmentation accuracy, particularly in critical pathological areas. We built a comprehensive diagnosis framework based on expert knowledge and clinical indicators, addressing multiple lesions like cervical disc herniation and spinal stenosis. Our proposed Relative Signal Change Index (RSCI) also enhanced T2 hyperintensity detection. The system outperformed CNN-based and Transformer-based models in both segmentation and diagnostic tasks, offering a reliable automated diagnostic solution. 

Our study has certain limitations. We only used T2-weighted MR images; incorporating multi-parametric MRI, X-ray, and CT data could enhance diagnostic accuracy. Future work will develop a multi-modality diagnosis system.  Furthermore, this study only included sagittal cervical MRI images. Future research will incorporate axial MRI images for segmentation and clinical indicators calculation. 



\section{References}

\bibliographystyle{ieeetr}
\bibliography{ref}


\end{document}